\documentclass[12pt,preprint]{aastex}

\newcommand{\etal}{{\it et al.}}

\begin{document}

\title{Terrestrial Planet Formation Around Individual Stars Within Binary Star Systems}

\author{Elisa V. Quintana} \affil{Space Science and Astrobiology
  Division 245-3, NASA Ames Research Center, Moffett Field, CA
  94035}\email{equintan@pollack.arc.nasa.gov}

\author{Fred C. Adams} \affil{Department of Physics, University of
  Michigan, Ann Arbor, MI 48109}

\author{Jack J. Lissauer} \affil{Space Science and Astrobiology
  Division 245-3, NASA Ames Research Center, Moffett Field, CA 94035}

\and

\author{John E. Chambers} \affil{Department of Terrestrial Magnetism,
  Carnegie Institution of Washington, Washington, DC, 20015}

\begin{abstract}

We calculate herein the late stages of
terrestrial planet accumulation around a solar type star that has a binary
companion with semimajor axis larger than the terrestrial planet
region.  We perform more than one hundred simulations to survey binary
parameter space and to account for sensitive dependence on initial
conditions in these dynamical systems. As expected, sufficiently wide
binaries leave the planet formation process largely unaffected. As a
rough approximation, binary stars with periastron $q_B > 10$ AU have
minimal effect on terrestrial planet formation within $\sim 2$ AU of
the primary, whereas binary stars with $q_B \la$ 5 AU restrict
terrestrial planet formation to within $\sim$ 1 AU of the primary star.  Given
the observed distribution of binary orbital elements for solar type
primaries, we estimate that about 40 -- 50 percent of the binary population
is wide enough to allow terrestrial planet formation to take place
unimpeded.  The large number of simulations allows for us to determine
the distribution of results --- the distribution of plausible
terrestrial planet systems --- for effectively equivalent starting
conditions.  We present (rough) distributions for the number of
planets, their masses, and their orbital elements.

\end{abstract}

\keywords{binary stars, circumstellar disks, planet formation,
  extrasolar planets}

\section{Introduction}
A binary star system is the most common result of the star formation
process, at least for solar-type stars, and the question of planet
formation in binaries is rapidly coming into focus.  At least 30 of
the first 170 extrasolar planets that have been detected are in
so-called S-type orbits, which encircle one component of a
main sequence binary/multiple star system (Eggenberger \etal\ 2004,
Raghavan \etal\ 2006, Butler \etal\ 2006). This sample includes 3
systems --- GJ 86 (Queloz \etal\ 2000), $\gamma$ Cephei (Hatzes
\etal\ 2003), and HD 41004 (Zucker \etal\ 2004) --- with stellar
semimajor axes of only $\sim$ 20 AU, well within the region spanned by
planets in our own Solar System.  Five of these 30 planets orbit one
member of a triple star system (Raghavan \etal\ 2006).  One example is
the planet HD 188753 Ab, which has a minimum mass of 1.14 times the
mass of Jupiter, $M_{J}$, and was detected in a 3.35 day S-type
orbit around a 1.06 $M_\odot$ star. A short-period binary star system
(with a total mass of 1.63 $M_\odot$) orbits this host-star/planet
system with a remarkably close periastron distance of $\sim$ 6 AU
(Konacki 2005).  The effects of the binary companion(s) on the
formation of these planets remain unclear, especially from a
theoretical perspective. The statistics of the observational data,
however, suggest that binarity has an effect on planetary masses and
orbits (Mazeh 2004, Eggenberger \etal\ 2004). The existence of
terrestrial-mass planets in $\it{main}$ $\it{sequence}$ binary star
systems remains observationally unconstrained.  The Kepler
Mission\footnote{see www.kepler.arc.nasa.gov}, set for launch in late
2008, has the potential to photometrically detect Earth-like planets
in both single and binary star systems, and will help provide such
constraints.


Some theoretical research on terrestrial planet formation in binary
star systems has already been carried out.  The growth stage from
km-sized planetesimals to the formation of Moon- to Mars-sized bodies
within a gaseous disk, via runaway and oligarchic growth, has been
numerically simulated with favorable results in the $\alpha$ Centauri
AB binary star system (Marzari and Scholl 2000) and the $\gamma$
Cephei binary/giant-planet system (Thebault \etal\ 2004).  In each
case, the combined effects of the stellar perturbations and gas drag
lead to periastron alignment of the planetesimal population, thereby
reducing the relative and collisional velocities and increasing the planetesimal
accretion efficiency within $\sim$ 2.5 AU of the central star.
Thebault \etal\ (2006) further examined this stage of planetesimal
growth in binary star systems with stellar separations $a_B \leq$ 50
AU and eccentricities $e_B \leq$ 0.9.  In addition to the damping of
secular perturbations on planetesimals in the presence of gas drag,
they found that the evolution is highly sensitive to the initial
size-distribution of the bodies in the disk, and binary systems with
$e_B \ga$ 0 can inhibit runaway accretion within a disk of
unequal-sized planetesimals.


The final stages of terrestrial planet formation - from planetary
embryos to planets - have been modeled for $\alpha$ Centauri AB, the
closest binary star system to the Sun (Barbieri \etal\ 2002, Quintana
\etal\ 2002, Turrini \etal\ 2005).  The G2 star $\alpha$ Cen A (1.1
M$_{\odot}$) and the K1 star $\alpha$ Cen B (0.91 M$_{\odot}$) are bound with a stellar separation of $a_B \sim$ 23.4 AU, a binary eccentricity
$e_B \sim$ 0.52, and have a stellar periastron $q_B \equiv a_B(1 - e_B)
\sim$ 11.2 AU.  Various disk mass distributions around $\alpha$ Cen A
were examined in Barbieri \etal\ (2002), and terrestrial planets
formed in their simulations with semimajor axes within $\sim$ 1.6 AU of $\alpha$ Cen A.  Quintana
\etal\ (2002) performed 33 simulations using virtually the same initial disk mass distribution (the `bimodal' model that is used in
the simulations presented in this article), and examined a range of
initial disk inclinations (0 to 60$^{\circ}$ and also 180$^{\circ}$)
relative to the binary orbital plane for a disk centered around
$\alpha$ Cen A, and performed a set of simulations with a disk centered around
$\alpha$ Cen B coplanar to the stellar orbit.  From 3 -- 5 terrestrial
planets formed around $\alpha$ Cen A in simulations for which the
midplane of the disk was initially inclined by 30$^{\circ}$ or less
relative to the stellar orbit (Quintana \etal\ 2002, Quintana 2004),
and from 2 -- 5 planets formed around $\alpha$ Cen B (Quintana 2003).
For comparison, growth from the same initial disk placed around the
Sun with neither giant planets nor a stellar companion perturbing the
system was simulated by Quintana \etal\ (2002).  Material remained
farther from the Sun, and accretion was much slower ($\sim$ 1 Gyr)
compared with the aforementioned $\alpha$ Cen AB integrations ($\sim$ 200 Myr) and
simulations of the Sun-Jupiter-Saturn system ($\sim$ 150 -- 200 Myr), all of
which began with virtually the same initial disk (Chambers 2001, Quintana \& Lissauer
2006).  These simulations show that giant and stellar companions not only truncate the
disk, but hasten the accretion process by stirring up the planetary
embryos to higher eccentricities and inclinations.  Raymond
\etal\ (2004) explored terrestrial planet growth around a Sun-like
star with a Jupiter-like planet in a wide range of initial
configurations (masses and orbits), and demonstrated that an eccentric
Jupiter clears out material in the asteroid region much faster than a
giant planet on a circular orbit.

The formation of planets in so-called P-type orbits which encircle
both stars of a binary star system has also been investigated.  Nelson
\& Papaloizou (2003) and Nelson (2003) studied the effects of a giant
protoplanet within a viscous circumbinary disk (with the total mass of
the stars and the protoplanet equal to 1 M$_{\odot}$), and found modes
of evolution in which the protoplanet can remain in stable orbits,
even around eccentric binary star systems.  The final stages of
terrestrial planet formation have been numerically simulated in P-type
circumbinary orbits around binary stars with a total mass of 1
M$_{\odot}$, semimajor axes 0.05 AU $\leq a_B \leq$ 0.4 AU, and binary
eccentricities $e_B \leq$ 0.8 (Quintana 2004, Lissauer \etal\ 2004,
Quintana \& Lissauer 2006).  These simulations began with the same
bimodal initial disk mass distribution as the model used in this
article, and the final planetary systems were statistically compared
to planetary systems formed in the Sun-Jupiter-Saturn system (which
used the same starting disk conditions).  Terrestrial planets similar
to those in the Solar System formed around binary stars with apastra
$Q_B \equiv a_B$(1 + $e_B$) $\la$ 0.2 AU, whereas simulations of
binaries with larger maximum separations resulted in fewer planets,
especially interior to 1 AU of the binary star center of mass.  Herein, we
only consider terrestrial planet formation around a single main
sequence star with a stellar companion, and aim to find similar
constraints on the stellar masses and orbits that allow the formation
of terrestrial planets similar to the Mercury-Venus-Earth-Mars system.


A related issue is that of the long term stability of planetary
systems in binaries (after planet formation has been completed).
Wiegert \& Holman (1997) examined test particles around one and both
stars in the $\alpha$ Cen AB system, and Holman \& Wiegert (1999)
explored the stability of orbits in and around (a wide range of)
binary star systems.  These two studies demonstrated that the
stability regions are dependent on four parameters: the binary
semimajor axis $a_B$, the binary eccentricity $e_B$, the inclination
of the disk relative to the stellar orbit, and the stellar mass ratio
$\mu \equiv M_C / (M_{\star} + M_C)$, where $M_{\star}$ is the mass of
the primary star and $M_C$ is the mass of the companion.  Dynamical
stability calculations of an Earth-like planet orbiting a Solar-mass
(M$_{\odot}$) star with an intermediate mass companion ($M_C$ = 0.001
-- 0.5 $M_\odot$) have shown that the regions of stability depend most
sensitively on the binary periastron $q_B \equiv a_B (1 - e_B)$, for a
given companion mass $M_C$ (David \etal\ 2003).  These stability
calculations, in conjunction with the observed distributions of binary
orbital parameters (Duquennoy \& Mayor 1991), indicate that $\sim$
50\% of all binary systems are wide enough so that Earth-like planets
can remain stable for the age of the Solar System.  Additionally, similar
calculations of bodies in circumbinary orbits (Quintana \& Lissauer 2006),
along with the observed binary period distribution (David \etal\ 2003), indicate that approximately 10\% of main sequence binary
stars are close enough to each other to allow the formation and long-term stability of Earth-like
planets.  One goal of this paper is to provide an analogous constraint
for planet $\it{formation}$ around individual stars in binary systems,
i.e., find the fraction of binaries that allow for an entire system of
terrestrial planets to form.  For the sake of definiteness, we
(somewhat arbitrarily) define a ``complete'' terrestrial planet system
to be one that extends out to at least the present orbit of Mars, $\sim$ 1.5 AU, which encompasses the inner terrestrial planets in the
Solar System and also the habitable zone of Sun-like stars (Kasting \etal\ 1993).


In this article, we present the results from a large survey ($\sim$
120 numerical simulations) on the effects of a stellar companion on
the final stages of terrestrial planet formation in S-type orbits
around one component of a main sequence binary star system.  We
examine stellar mass ratios of $\mu$ = 1/3, 1/2, and 2/3 (for stars
with masses of either 0.5 $M_\odot$ or 1 $M_\odot$), and the stellar
orbital parameters ($a_B$, $e_B$) are varied such that the systems
take on periastron values of $q_B$ = 5, 7.5, or 10 AU.  Multiple
simulations, usually 3 to 10, are performed for each binary star
system with slight changes in the starting states in order to explore
the sensitivity of the planet formation process to the initial
conditions. In one case, we performed 30 integrations of a single
system (with $q_B$ = 7.5 AU, $a_B$ = 10 AU, and $e_B$ = 0.25) to
produce a large distribution of final planetary systems in order to
quantify this chaotic effect.  This paper is organized as follows.
The orbital stability of test particles in S-type orbits in binary
star systems is discussed in \S 2.  The numerical model and initial
conditions are presented in \S 3, and results of our planetary
accretion simulations are presented and discussed in \S 4.  We
conclude in \S 5 with a summary of our results and their implications.

\section{Orbital Stability}
Prior to examining planetary accretion in binaries, we briefly
consider the stability of test particles in the binary systems of
interest. In approximate terms, if a binary is wide enough so that
test particles in orbit about the primary remain largely unperturbed
over a time span of $\sim 100$ Myr (characteristic of terrestrial
planet formation), one would not expect the binary to have a large
impact on planet formation.  The regions of orbital stability of
planets in S-type orbits have been investigated previously for a wide
range of binary star systems (Holman \& Wiegert 1999, David
\etal\ 2003).  Holman \& Wiegert (1999) performed a dynamical analysis
for stellar mass ratios in the range 0.1 $\le \mu \le$ 0.9, and for
binary eccentricities 0 $\le e_B \le$ 0.8. In their study, test
particles were placed in the binary orbital plane between 0.02 $a_B$
and 0.5 $a_B$ at eight equally spaced longitudes per semimajor axis,
and the system was evolved forward in time for 10$^4$ binary periods.
The outermost stable orbit to the primary star for which all eight
particles survived (which they refer to as the `critical semimajor
axis', $a_c$) was determined for each system.

We have performed analogous test particle simulations for the binary
star systems examined in this article, which include equal mass ($\mu$
= 0.5) binary stars (with eccentricities 0 $\leq e_B \leq$ 0.875), and
binary systems with $\mu$ = 1/3 or 2/3 (where $e_B$ = 0.25 in each
case).  Test particles were placed between 0.02 $a_B$ and 0.5 $a_B$
(with increments of 0.01 $a_B$) at eight equally spaced longitudes per
semimajor axis.  We use a `wide-binary' symplectic algorithm developed
by Chambers \etal\ (2002) for these and subsequent accretion
simulations.  This algorithm is based on the mapping method of Wisdom
\& Holman (1991), and was designed to calculate the evolution of
particles/bodies around a central star with a massive companion
perturbing the primary star/disk system.  Each system was followed for
10$^6$ binary orbits, and the critical semimajor axis (for which all
eight particles remained in the system) was calculated and is given in
Table 1\footnote{We note that David et al. (2003) performed similar
  calculations using both a symplectic code and a Bulirsh-Stoer
  integration scheme; both numerical algorithms produced the same
  results, i.e., the same distribution of ejection times}.  Note that
some of the systems used here have values of $e_B$ in between those
examined in Holman \& Weigert (1999), and upon extrapolation the
results presented in Table 1 are consistent with their study.

\section{Numerical Model and Initial Conditions}
Each simulation begins with a disk of planetesimals and planetary
embryos centered around what we will refer to as the primary star
(even though for most simulations the stars have equal masses, and in
one set of runs the planetesimals and embryos encircle the lower mass
star), with the midplane of the disk coplanar to the stellar orbit,
and the binary companion in an orbit exterior to the terrestrial
planet region.  These simulations use an initial disk composed of
planetesimals and embryos with masses and orbits virtually identical
to those used in previous simulations of terrestrial planet formation
in the Sun-Jupiter-Saturn System, in particular, the model of Chambers
(2001) that was most successful in reproducing the terrestrial planets
in our Solar System.  Although this formulation is not complete, nor
definitive, it provides a model that reproduces our terrestrial planet
system (albeit with somewhat larger eccentricities) and can thus be
used as a reference point.  By performing a large ensemble of
analogous simulations in binary systems with varying orbital elements,
we can assess the effects of binarity on the planet formation process.

The disk is composed of Moon- to Mars-sized rocky bodies which are
assumed to have already formed from a disk of gas and dust.  In
particular, 14 planetary embryos (each 0.0933 times the mass of the
Earth, M$_{\oplus}$) compose half of the disk mass, while the
remaining mass is distributed equally among 140 planetesimals (each
with a mass of 0.00933 M$_{\oplus}$), providing a total disk mass of
$\sim$ 2.6 M$_{\oplus}$.  The initial radial extent of the disk is
between 0.36 AU -- 2.05 AU of the primary star, and the radius of each
body is calculated assuming a material density of 3 g cm$^{-3}$.  The
protoplanets began with initial eccentricities of $e \leq$ 0.01,
inclinations $i \leq$ 0.5$^{\circ}$, and specific initial orbital
elements were chosen at random from specified ranges; the same set of
randomly selected values was used for all simulations.  


The `wide-binary' algorithm that we use follows the evolution of each body
in the disk subject to gravitational perturbations from both stars, and to gravitational interactions and completely inelastic
collisions with other bodies in the disk (Chambers \etal\ 2002).
Material that is not accreted onto growing planets may be lost from
the system by either orbiting too close to the central star (for the
simulations presented herein, planetesimals/embryos are removed when
their periastron $\leq$ 0.1 AU), or if it is ejected from the system
(if its trajectory exceeds a distance comparable to the binary
semimajor axis).  A time-step of 7 days is used, and the evolution of
each system is followed for 200 -- 500 Myr.  For a discussion on the
validity of using a 7-day time-step for this accretion disk, see
Section 2.2 of Quintana \& Lissauer (2006).

Many of our simulations began with binary stars on highly eccentric
orbits, which would perturb disk material into more elliptical orbits
than is considered in our chosen initial conditions.  One might worry
that this mismatch in initial conditions could produce a systematic
bias in our results.  This issue was investigated in Appendix B of
Quintana \& Lissauer (2006), in which the standard bimodal disk (as
previously described) was centered around close binary
stars.  It was shown that beginning the protoplanets with the
appropriate nested elliptical orbits, rather than in nearly circular
orbits, did not impact the resulting planetary systems in a
statistically significant manner.  In other words, the systematic
effect from the differing initial eccentricities of bodies in each
disk on the final planet systems that form (in otherwise identical
systems) were comparable to or smaller than the effects resulting from
chaos.  These results were also consistent with test particle
simulations in eccentric binary star systems performed by Pichardo
\etal\ (2005).

We examine binary star systems with stellar mass ratios $\mu \equiv
M_C / (M_{\star} + M_C)$ = 1/3, 1/2, or 2/3, where $M_{\star}$ is the
mass of the primary star and $M_C$ is the mass of the companion.
Table 2 displays the stellar parameters used in our accretion
simulations, grouped into four sets according to the stellar masses
involved.  The majority of our simulations begin with equal mass stars
($\mu$ = 1/2) of either $M_{\star}$ = $M_C$ = 0.5 M$_{\odot}$ (Set A)
or $M_{\star}$ = $M_C$ = 1 M$_{\odot}$ (Set B).  In the simulations of
Set C, the primary star (the one the disk is centered around) is more
massive, $M_{\star}$ = 1 M$_{\odot}$ and $M_C$ = 0.5 M$_{\odot}$
($\mu$ = 1/3).  In Set D, the primary star is smaller, $M_{\star}$ = 0.5
M$_{\odot}$ and $M_C$ = 1 M$_{\odot}$ ($\mu$ = 2/3).  The stellar
semimajor axis $a_B$ and binary eccentricity $e_B$ are varied such
that the binary periastron takes one of
the three values $q_B$ = 5, 7.5, or 10 AU (see columns 2 -- 4 of
Table 2).  Note that binary systems with much wider periastra would
have little effect on terrestrial planet formation, whereas systems
with smaller periastra would completely destroy the initial disk of
planetesimals.  The binary stars are separated by $a_B$ = 10, 13$\frac{1}{3}$, 20, or 40 AU, and the eccentricities are varied
in the range 0 $\leq e_B \leq$ 0.875.  The fifth column of Table 2
gives the outermost stable orbit for each system, $a_{c}$ (expressed
here in AU), as described in the previous section.  The largest
semimajor axis for which particles can be stable in any of the systems
that we explore is 2.6 AU (Table 2), we therefore omit giant planets
analogous to those in the Solar System (which orbit beyond 5 AU) in
our integrations.

For our accretion simulations, our exploration of parameter space has
two coupled goals.  On one hand, we want to determine the effects of
the binary orbital elements on the final terrestrial planet systems
produced. On the other hand, for a given binary configuration, we want
to explore the distribution of possible resulting planetary systems
(where the results must be described in terms of a distribution due to
the sensitive dependence on the initial conditions). Toward these
ends, we have performed from 3 -- 30 integrations (last column in
Table 2) for each binary star configuration ($\mu$, $a_B$, and $e_B$)
considered herein, with small differences in the initial conditions: a
single planetesimal near 0.5, 1, or 1.5 AU is moved forward along its
orbit by a small amount (1 -- 9 meters) prior to the integration.
Ideally, of course, one would perform larger numbers of integrations
to more fully sample the distributions of results, but computer
resources limit our sample size.


\section{Planetary Accretion Simulations}

This section presents the results of our numerical simulations. We
first consider the chaotic nature of the integrations (\S 4.1) and
discuss how sensitive dependence on initial conditions affects the
nature of our results. In \S 4.2, we present a broad overview of our
simulations by showing typical sequences of evolution and general
trends emerging in the results.  We focus on the properties of final
planetary systems in \S 4.3, where such properties are described in
terms of distributions.

\subsection{Chaotic Effects} 
Before comparing the results of terrestrial planet growth within
various binary star systems, it is worthwhile to take a closer look at
the stochastic nature of these $N$-body simulations, and keep in mind
throughout this study: {\it ``Long term integrations are not
  ephemerides, but probes of qualitative and statistical properties of
  the orbits \dots''} (Saha \etal\ 1997).  To demonstrate the
sensitive dependence on the initial conditions, Figure 1 shows the
evolution of two nearly identical simulations (with $a_B$ = 10 AU,
$e_B$ = 0.25, and $M_{\star}$ = $M_C$ = 1 M$_{\odot}$) that differ
only in a 1 meter shift of one planetesimal near 1 AU prior to the
integration.  Each panel shows the eccentricity of each body in the
disk as a function of semimajor axis at the specified time, and the
radius of each symbol is proportional to the radius of the body that
it represents.  The first two columns show the temporal evolution from
each simulation, and the third column plots the discrepancy between
the two systems at the corresponding simulation time:
planetesimals/embryos from columns one and two with orbits which
diverged in $(a, e)$ space by more than 0.001 are shown in red and in
blue, respectively.  At 100 years into the simulation, the two disks
shown in Figure 1 are virtually identical.  Although there are no
ejections nor collisions within the first 500 years of either
simulation, the bodies in the disk are dynamically excited, especially
towards the outer edge of the disk, and begin to exhibit chaotic
behavior.  The general behavior of the disk is similar among the two
systems, yet the stochastic nature leads to remarkably different final
planetary systems.

Figures 2a and 2b each show the orbital elements of a single
planetesimal from the two simulations presented in Figure 1.  The
semimajor axis ($a_p$), eccentricity ($e_p$), and inclination relative to
the binary orbital plane ($i_p$) are shown as a function of time for the
first 1000 years of the integrations.  In Figure 2a, the two
planetesimals (shown in red and in blue) began with the same semimajor
axis near 1 AU, and differed only by a 1 meter shift in their mean
anomaly.  The two planetesimals shown in Figure 2b, from the same two
simulations as the planetesimals in Figure 2a, began with identical
initial orbits near 0.95 AU.  The values of each element oscillate
with time due to the combined effect of the stellar and planetary
perturbations, and the orbits of two planetesimals clearly diverge
from one another on timescales longer than their short period
oscillations.  The divergence between the simulations is mainly due to discrete events, i.e. close encounters between planetesimals/embryos, and the Lyapunov time
for these $N$-body systems (where $N$ = 156) is of order $10^2$ years.
In Figure 2a, the planetesimal shown in red was ultimately ejected at
6.8 Myr into the first integration, while the planetesimal shown in blue fell
to within 0.1 AU of the central star at 10.6 Myr into the second
integration.  In Figure 2b, the planetesimal shown in red collided
with another planetesimal at 1.4 Myr, and this more massive body was later accreted by an embryo at 58 Myr.  The planetesimal shown in blue in
Figure 2b, however, was swept up by a larger embryo at 47.6 Myr into
the simulation.  Although small differences in initial conditions
(among an otherwise identical system) can ultimately lead to the
formation of entirely different terrestrial planet systems (number,
masses, and orbits, etc.), the early evolution of the disk is similar
among systems with the same binary star parameters, and there are
clear trends in the final planetary systems that form (as presented
and discussed in the next two subsections).

\subsection{Overview of Results}

Table 3 displays, for each binary star configuration, a set of
statistics developed to help quantify the final planetary systems
formed.  Each ($\mu$, $a_B$, $e_B$) configuration is listed in column
1, and the number of runs performed for each configuration are given
in column 2.  Columns 3 -- 13 are defined as follows (see Chambers
2001, Quintana \etal\ 2002, and Quintana \& Lissauer 2006 for
mathematical descriptions of most of these statistics).  Note that,
with the exception of column 5, only the average value of each
statistic is given for each set.

\begin{description}
\item[(3)] The average number of planets, $N_p$, at least as massive
  as the planet Mercury ($\sim$ 0.06 M$_{\oplus}$), that form in the
  system.  Note that each of the 14 planetary embryos in the initial
  disk satisfy this mass requirement, as do bodies consisting of at
  least 7 planetesimals.

\item[(4)] The average number of minor planets, $N_m$, less massive than the
  planet Mercury that remain in the final system.

\item[(5)] The maximum semimajor axis of the final planets, $a_{p_{max}}$.

\item[(6)] The average value of the ratio of the maximum semimajor
  axis to the outermost stable orbit of the system,
  $a_{p_{max}}$/$a_{c}$.

\item[(7)] The maximum apastron of the final planets, $Q_{p} \equiv$ $a_p$(1
  + $e_p$). 

\item[(8)] The average value of the ratio of the maximum apastron $Q_{p}$ to the binary
  periastron $q_B \equiv$ $a_B$(1 -- $e_B$).

\item[(9)] The fraction of the final mass in the largest planet,
  $S_m$.

\item[(10)] The percentage of the initial mass that was lost,
  $m_{l_\star}$, due to close encounters ($\leq$ 0.1 AU) with the primary
  star.

\item[(11)] The percentage of the initial mass that was ejected from
  the system without passing within 0.1 AU of the primary star,
  $m_{l_\infty}$.

\item[(12)] The total mechanical (kinetic + potential) energy per unit
  mass for the planets remaining at the end of a simulation, $E$,
  normalized by $E_{0}$, the energy per unit mass of the disk bodies
  at the beginning of the integration.

\item[(13)] The angular momentum per unit mass of the final planets,
  $L$, normalized by $L_0$, the angular momentum per unit mass of the
  initial system.

\end{description}

For comparison, following the results from Sets A - D are analogous
statistics for the following systems: the terrestrial planets
Mercury-Venus-Earth-Mars in the Solar System (`MVEM', of which only
three are actual observables); the averaged values for 31 accretion
simulations in the Sun-Jupiter-Saturn system (`SJS$\_{\rm{ave}}$',
Chambers 2001, Quintana \& Lissauer 2006); the averaged values from a
set of accretion simulations around the Sun with neither giant planets
nor a stellar companion perturbing the system (`Sun$\_{\rm{ave}}$',
Quintana \etal\ 2002); the averaged values for the planets formed
within 2 AU of the Sun-only simulations (`Sun$\_{\rm{ave}}$ ($a <$ 2
AU)', Quintana \etal\ 2002); and the averaged values for the planetary
systems formed around $\alpha$ Cen A in simulations for which the disk
began coplanar ($i =$ 0$^{\circ}$) to the $\alpha$ Cen AB binary
orbital plane (`$\alpha$ Cen ($i =$ 0$^{\circ}$)', Quintana
\etal\ 2002).

Table 3 shows both the striking uniformity of (many of) these
statistical measures, in spite of the great diversity of stellar
configurations under consideration, and the systematic variation of
the output measures with initial conditions.  For example, column 8
shows that terrestrial planet systems have a ``nearly constant'' size,
measured as a fraction of binary periastron. On one hand, this size
ratio lies in the range 0.12 to 0.23 for each set of simulations.  On
the other hand, this size ratio clearly varies systematically with
variations in the class of initial conditions. Of course, the ratio
varies from system to system within the same class of starting
conditions.  Similar behavior occurs for all of these statistical
measures: they all have well-defined characteristic values for the
entire (diverse) set of systems under consideration, they all vary
systematically with the class of initial conditions, and they all vary
stochastically from case to case for systems within the same
(effectively equivalent) class of initial conditions.

A visual comparison of the effects from various binary stars on the
evolution of the disk of planetary embryos is given in Figures 3 -- 6,
each of which provides a side-by-side view of three simulations that
have two of the four stellar parameters ($\mu$, $a_B$, $e_B$, $q_B$)
in common.  Figure 3 shows the evolution of three systems from Set A,
each with $\mu$ = 0.5 ($M_{\star}$ = $M_C$ = 0.5 M$_{\odot}$) and
$a_B$ = 20 AU.  The simulations differ only in the stellar
eccentricities (and therefore periastra): $e_B$ = 0.75 and $q_B$ = 5
AU (first column), $e_B$ = 0.625 and $q_B$ = 7.5 AU (middle column),
and $e_B$ = 0.5 and $q_B$ = 10 AU (third column).  The temporal
evolution of a single simulation is shown with black solid circles
which represent (and are proportional to) the bodies in the disk.
Additionally, the last row shows the final planetary systems that
formed in 2 other realizations of the same system (shown with gray and
open circles in Figures 3 -- 6).  In the first two columns of Figure
3, binary systems with $q_B$ = 5 AU and 7 AU, the stellar companion
truncates the disk to within 2 AU early in the simulation.  The high
binary star eccentricities stir up both the planetesimals and embryos,
especially towards the outer edge of the disk.  Among the three
realizations of the $q_B$ = 5 AU system, 40\% of the initial mass on
average was perturbed to within 0.1 AU of the central star, and an
average of 31\% of the initial disk mass was ejected from the system
(see columns 10 and 11 of Table 3).  The remaining mass was accreted
into 1 or 2 final planets, with semimajor axes $a_p \la$ 0.7 AU, on
timescales of $\sim$ 20 -- 50 Myr.

The middle column of Figure 3 shows the $q_B$ = 7.5 AU ($e_B$ = 0.625)
system, in which an average of $\sim$ 39\% of the material was lost
into the central star, and only 9\% of the initial mass was ejected
from the system.  Most of the accretion was complete by $\sim$ 100
Myr, resulting in the formation of 2 -- 3 planets with $a_p \la$ 1.3
AU (the largest apastron of any planet in the set was $Q_p$ = 1.4 AU).
In the $q_B$ = 10 AU (and $e_B$ = 0.5) simulations, shown in the third
column of Figure 3, the eccentricities of the bodies in the disk are
perturbed to high values, while the disk extends to beyond 2 AU within
the first 20 Myr.  In these runs, an average of $\sim$ 31\% of the
initial mass is lost into the primary star, $\sim$ 2\% is ejected, and
2 or 3 planets remain within $Q_p \la$ 1.8 AU.  The fraction of mass
that composes the largest planet, $S_m$ (column 9 of Table 3),
decreases with increasing $q_B$ (from 0.95 -- 0.53 in the systems
shown in Figure 3), and is anti-correlated with the number of final
planets that form.  The magnitude of the specific energy of the $q_B$
= 5, 7.5, and 10 AU systems increase by 76\%, 43\%, and 31\%,
respectively, while the specific angular momentum of these systems
decrease by 29\%, 19\%, and 13\% from their original values (columns
12 and 13 of Table 3).

Figure 4 shows three simulations from Set A with the same stellar
periastron $q_B$ = 7.5 AU, but with differing values of ($a_B$,
$e_B$).  In the first column, in which $a_B$ = 10 AU and $e_B$ = 0.25,
the planetesimals are highly excited, yet most of the larger embryos
in the disk remain with $e_p$ $\la$ 0.1 within the first few million
years.  In the middle and final column, in which the semimajor axis
and the eccentricity ($a_B$, $e_B$) are both increased to (20 AU, 0.625)
and (40 AU, 0.8125), respectively, the more massive embryos are also
excited to high values of $e_p$.  The total mass that was lost is
comparable among the three simulations, and also among the additional realizations of the three systems shown in Figure 4: 49\% on average when
$a_B$ = 10 AU, 48\% for $a_B$ = 20 AU, and 52\% for $a_B$ = 40 AU.
The more eccentric binary stars tend to stir up the outer edge of the
disk such that most of the mass that is lost is perturbed
to within 0.1 AU of the primary star; approximately 82\% and 65\% of the
mass that was lost in the second ($a_B$ = 20 AU, $e_B$ = 0.625) and third ($a_B$ = 40 AU,
$e_B$ = 0.8125) simulations was perturbed into the central star, whereas 80\% of the
mass lost in the simulation shown in the first panel ($a_B$ = 10
AU, $e_B$ = 0.25) was ejected from the system.  Although the
total change in energy and angular momentum of each system is
comparable, fewer planets form, with the outermost planet closer to
the central star, for binary stars on more eccentric orbits.  The
effect of a more eccentric binary star system for a given $q_B$ is
typically a more diverse set of planetary systems, which was also the
case for terrestrial planet formation within a circumbinary disk
surrounding highly eccentric close binary stars (but in that case the
apastron value was the pertinent parameter, Quintana \& Lissauer
2006).

Three simulations from Set B ($M_{\star}$ = $M_C$ = 1.0 M$_{\odot}$)
with $a_B$ = 10 AU (and differing $e_B$ and $q_B$) are shown in Figure
5.  Although the binary stars in Set B are each twice as massive as
those in Set A, the mass ratio is the same ($\mu$ = 0.5) and for a
given ($a_B$, $e_B$), or $q_B$, the simulations display similar trends
(Figure 5) as those from Set A (Figure 3).  Note that only three final
planetary systems (from three realizations chosen at random) are shown in the last row
of each column in Figures 5 and 6, even though from 10 -- 30
integrations were performed, in order to demonstrate with clarity the diversity of
final planets among each set.  For simulations with $q_B$ = 5 AU
($a_B$ = 10 AU and $e_B$ = 0.5), an average of 36\% of the initial
disk mass was lost to the central star, and 29\% was ejected from the
system.  In the 10 runs of this system, from 1 -- 2 terrestrial-mass
planets formed with $Q_p <$ 1 AU of the central star.  We performed
an ensemble of 30 integrations for a system with $q_B$ = 7.5 AU, $a_B$
= 10 AU, and $e_B$ = 0.25 system, three of which are shown in the
middle column of Figure 5.  The average percentage of mass that falls
within 0.1 AU of the star is 10\%, consistent with the analogous runs
from Set A, whereas the mass perturbed out of the system is somewhat
less, 30\% on average.  Within the first 50 Myr, from 2 -- 5 planets
at least as massive as Mercury have accreted, with 2 -- 5
planetesimals remaining in each system on highly eccentric orbits.
From 1 -- 4 final terrestrial planets formed (with an average of 2.8)
with $Q_p \la$ 1.8 AU of the primary star.  The somewhat less mass
that is lost in Set B compared with an identical orbital parameter
system from Set A can be expected, as although stellar perturbations
are directly scaled, planetary perturbations are relatively less
significant in Set B since the mass ratio of the planet to the star is
smaller, which results in less internal excitation (see Appendix C of
Quintana \& Lissauer 2006 for a discussion of scaling planetary
accretion simulations).  In simulations from Set B with larger
periastra of $q_B$ = 10 AU ($a_B$ = 10 AU, $e_B$ = 0), such as those
shown in the third column of Figure 5, from 2 -- 5 planets form with
semimajor axes that encompass the full range of the initial disk, and
even beyond 2 AU (one planet remained with $Q_p$ = 2.78), with 26\% of
the initial disk mass lost via ejection.  The disk is slightly more
truncated in simulations from Set B with $q_B$ = 10 AU, $a_B$ = 40 AU
and $e_B$ = 0.75 (not shown), and from 1 -- 4 planets accreted with
$Q_p <$ 2 AU by the end of these integrations.

Figure 6 presents the evolution of three simulations with the same
orbital parameters, $a_B$ = 10 AU and $e_B$ = 0.25 ($q_B$ = 7.5 AU),
but examines the effect of different stellar mass ratios.  In the
first column, the central star has a mass of 0.5 M$_{\odot}$, and a 1
M$_{\odot}$ star perturbs the disk.  Among this set, an average of
72\% of the initial disk mass is cleared out (all of it into
interstellar space), and the remaining mass is accreted into 2 -- 4
planets with $Q_p \la$ 1.1 AU.  The middle column shows three
additional simulations from the $q_B$ = 7.5 AU ($a_B$ = 10 AU, $e_B$ =
0.25) system from Set B that is shown in the middle column of Figure
5, in which $\sim$ 40\% of the initial mass is lost on average. The
final column shows a simulation with the disk centered around a 1
M$_{\odot}$ star with a 0.5 M$_{\odot}$ stellar companion.  In this
set of runs with $\mu$ = 1/3, an average of 26\% of the initial disk
mass is lost (most of it is perturbed to within 0.1 AU the central
star), and 3 or 4 planets accrete with $Q_p \la$ 1.8 AU in each
simulation.  This configuration statistically has the same effect on
the disk and the resulting planets as Jupiter and Saturn do in
analogous simulations of the Solar System (Chambers \etal\ 2001,
Quintana \& Lissauer 2006).  Note that the temporal evolution of most
individual simulations from Sets A and B are presented in Quintana
(2004).

\subsection{Final Planetary Systems}

As described in the previous section, the stellar mass ratio and the
periastron distance strongly influence where terrestrial planets can
form in binary star systems.  The effect of $q_B$ on the distribution
of final planetary system parameters (i.e., number, masses, etc.) is
further explored in this section.  Figure 7 displays the distribution
of planetary eccentricities and semimajor axes for all of the final planets formed in equal-mass binary systems, with symbol sizes proportional to planet sizes, with $q_B$ =
5 AU (top panel), 7.5 AU (middle panel), and 10 AU (lower panel).
Systems with $M_{\star}$ = $M_C$ = 0.5 M$_{\odot}$ (Set A) are
represented in red, while systems with $M_{\star}$ = $M_C$ = 1.0
M$_{\odot}$ (Set B) are shown in blue.  Although the initial
planetesimals/embryos in the disk begin with semimajor axes that
extend out to 2 AU, all of the planets formed in systems with $q_B$ =
5 AU remained with $a_p <$ 0.9 AU of the primary star regardless of
the ($a_B$, $e_B$) values.  The simulations of systems with $q_B$ =
7.5 AU resulted in the formation of terrestrial-mass planets within
the current orbit of Mars ($\sim$ 1.5 AU).  The final planets
formed in systems with $q_B$ = 10 AU have a wider range of both $a_p$
(out to 2.2 AU) and $e_p$ (up to 0.45).  The results from Set A and
Set B are comparable for a given $q_B$, consistent with the stability
constraints described in Section 2.  Note the pile-up, however, of
planetesimals in the inner region of the disk in systems with
$M_{\star}$ = $M_C$ = 1.0 M$_{\odot}$.  The innermost initial
planetesimal often survives intact for Set B (in which the ratio of the disk mass to
star mass is smaller), but not in Set A.

The semimajor axis of the outermost planet can be used as a measure of
the size of the terrestial planet system.  Figure 8 shows the
distribution of the semimajor axis of the outermost final planet,
$a_{p_{max}}$, formed in each simulation for systems with $q_B$ = 5 AU
(top panel), 7.5 AU, (middle panel), and 10 AU (lower panel).  Note
that twice as many integrations have been performed in Set B than in
Set A.  Figure 8 shows a clear trend: as the binary periastron
increases, the distribution of semimajor axes of the outermost planet
becomes wider and its expectation value shifts to larger values.  This
trend of a larger distribution in semimajor axis for increasing $q_B$
is also demonstrated in Figure 9, for which the outermost semimajor
axis of the final planets is plotted directly as a function of $q_B$.
Both the expectation value and the width of each distribution clearly
grow (in a nearly linear fashion) with increasing values of binary
periastron $q_B$.

The distributions of the total number of final planets formed are
shown in Figure 10 for simulations with $q_B$ = 5 AU (top panel), 7.5
AU, (middle panel), and 10 AU (lower panel).  In general, a smaller
binary periastron results in a larger percentage of mass loss, and a
smaller number of final planets.  From 1 -- 3 planets formed in all
systems with $q_B$ = 5 AU, 1 -- 5 planets remained in systems with
$q_B$ = 7.5 AU, and 1 -- 6 planets formed in all systems with $q_B$ =
10 AU.  The range in the number of possible planets thus grows with
increasing binary periastron; similarly, the average number of planets
formed in the simulations is an increasing function of $q_B$.  In the
$q_B$ = 7.5 AU set with equal mass stars of 1 M$_{\odot}$, shown in
blue, an average of 2.8 planets formed in the distribution of our
largest set of 30 integrations.  The distribution extends farther out
if the perturbing star is smaller relative to the central star for a given
stellar mass ratio, and slightly farther out when the stars are more
massive relative to the disk.

Figure 11 shows the distribution of final planetary masses (in units
of the Earth's mass, M$_{\oplus}$) formed in systems with $q_B$ = 5 AU
(top panel), 7.5 AU, (middle panel), and 10 AU (lower panel).  The
median mass of the final planets doesn't depend greatly on $q_B$.
This result suggests that planet formation remains quite efficient in
the stable regions, but that the size of the stable region shrinks as
$q_B$ gets smaller. This trend is consistent with the decline in the
number of planets seen in Figure 10.  When the periastron value becomes
as small as 5 AU, planets only form within 1 AU, and the mass
distribution tilts toward $m_p < M_{\oplus}$, i.e., the formation of
Earth-like planets is compromised.  Figure 12 displays the
distribution of eccentricities for the planets that are more massive
than the planet Mars that remain in systems with $q_B$ = 5 AU (top
panel), $q_B$ = 7.5 AU (middle panel), and $q_B$ = 10 AU (bottom
panel).  The distribution of eccentricities reaches relatively high
values in several simulations (up to $\sim$ 0.45 in the system with
the largest periastron).  However, the majority of planets orbit on
nearly circular ($e \la$ 0.1) orbits.

\section{Conclusions}
This paper studies the late stages of terrestrial planet formation in
binary star systems. These dynamical systems -- disks of planetesimals
orbiting in the potential of a binary star -- are chaotic and display
extremely sensitive dependence on their initial conditions. In the
present context, as for analogous models of terrestrial planet
formation around a single star (Chambers \etal\ 2002, Quintana \&
Lissauer 2006), if one of the planetesimals is moved only one meter
forward along its initial orbit, the difference in the final number,
masses, and/or orbits of terrestrial planets that form can be
substantial (see Figure 1).  Effectively equivalent starting states thus
lead to different final states. As a result, the numerical simulation
of accretion from any particular initial configuration cannot be
described in terms of a single outcome, but rather must be considered
as a distribution of outcomes.  To account for this complication, we
performed multiple realizations with effectively equivalent starting
conditions for all of the binary systems considered herein.
Fortunately, even though the systems are chaotic and do not have a
single outcome, the distributions of outcomes are well-defined.
Further, these distributions show clear trends with varying properties
of the binary star systems and can be used to study the effects
of binaries on the planet formation process.

Our exploration of parameter space shows how binary orbital parameters
affect terrestrial planet formation. We find that the presence of a
binary companion of order 10 AU away acts to limit the number of
terrestrial planets formed and the spatial extent of the terrestrial
planet region, as shown by Figures 3 -- 12.  To leading order, the
periastron value $q_B$ is the most important parameter in determining
binary effects on planetary outcomes (more predictive than $a_B$ or
$e_B$ alone).  In our ensemble of 108 wide binary simulations that
began with equal mass stars, from 1 -- 6 planets formed with semimajor
axes $\la$ 2.2 AU of the central star in binary systems with $q_B$ =
10 AU, from 1 -- 5 planets formed within 1.7 AU for systems with $q_B$
= 7.5 AU, and from 1 -- 3 planets formed within 0.9 AU when $q_B$ = 5
AU.  Nonetheless, for a given binary perisatron $q_B$, fewer planets
tend to form in binary systems with larger values of ($a_B$, $e_B$)
(see Figure 4).

Binary companions also limit the extent of the terrestrial planet
region in nascent solar systems. As shown in Figures 3, 5, and 7 -- 9,
wider binaries allow for spatially larger systems of terrestrial planets. Once
again, binary periastron is the most important variable in determining
the extent of the final system of terrestrial planets (as measured by
the semimajor axis of the outermost planet).  However, for a given
periastron, the sizes of the terrestrial planet systems show a wide
distribution.  In these simulations, the initial disk of planetesimals
extends out to 2 AU, so we do not expect terrestrial planets to form
much beyond this radius. For binary periastron $q_B$ = 10 AU, the
semimajor axis of the outermost planet typically lies near 2 AU, i.e.,
the system explores the entire available parameter space for planet
formation. Since these results were obtained with equal mass stars
(including those with $M$ = 1.0 M$_{\odot}$), we conclude that the
constraint $q_B \ga 10$ AU is sufficient for binaries to leave
terrestrial planet systems unperturbed.  With smaller binary
periastron values, the resulting extent of the terrestrial planet
region is diminished. When binary periastron decreases to 5 AU, the
typical system extends only out to $a_p \sim$ 0.75 AU and no system
has a planet with semimajor axis beyond 0.9 AU (but note that we did
not perform simulations with $q_B =$ 5 AU and small $e_B$).

While the number of forming planets and their range of orbits is
restricted by binary companions, the masses and eccentricities of
those planets are much less affected. Both Figures 7 and 12 suggest
that planet eccentricities tend to arise from the same distribution
(with a broad peak near $e_P = 0.08$ and a tail toward higher values)
for all of the systems considered in this study. The distribution of
planet masses is nearly independent of binary periastron (see Figure
11), although the wider binaries allow for a few slightly more massive
terrestrial planets to form. Finally, we note that the time scales
required for terrestrial planet formation in these systems lie in the
range 50 -- 200 Myr, consistent with previous findings (e.g., Chambers
2001, Quintana \etal\ 2002, Quintana \& Lissauer 2006), and largely
independent of the binary properties. This result is not unexpected,
as the clock for the accumulation of planetesimals is set by their
orbit time and masses (Safranov 1969), and not by the binary orbital
period.

Whitmire et al. (1998) have analyzed the effects of perturbations by a
binary companion on planetesimals during the earlier stages of
planetary growth.  Assuming that collisions at velocities $>$ 100 m/s
disrupt planetesimals, they find that if two 1 M$_\odot$ stars have
periastron $q_B <$ 16 AU, then planetary growth at 1 AU is inhibited.
This criterion is more limiting than the results of this paper
suggest.  But note that Whitmire et al.'s model does not include gas.
Perturbations by a gaseous disk can align planetesimal orbits,
reducing collision velocities and thereby allowing growth to proceed
and produce bodies of the sizes that we use as initial conditions over
a larger range of semimajor axis (Kortenkamp \& Wetherill 2000).

Our work has important implications regarding the question of what
fraction of stars might harbor terrestrial planetary systems.  The
majority of solar-type stars live in binary systems, and binary
companions can disrupt both the formation of terrestrial planets and
their long term prospects for stability.  Approximately half of the
known binary systems are wide enough (in this context, having
sufficiently large values of periastron) so that Earth-like planets
can remain stable over the entire 4.6 Gyr age of our Solar System
(David \etal\ 2003, Fatuzzo \etal\ 2006). For the system to be stable
out to the distance of Mars's orbit, the binary periastron $q_B$ must
be greater than about 7 AU, and about half of the observed binaries
have $q_B > 7$ AU.  Our work on the formation of terrestrial planets
shows similar trends.  When the periastron of the binary is larger
than about $q_B$ = 10 AU, even for the case of equal mass stars,
terrestrial planets can form over essentially the entire range of
orbits allowed for single stars (out to the edge of the initial
planetessimal disk at 2 AU). When periastron $q_B < 10$ AU, however,
the distributions of planetary orbital parameters are strongly
affected by the presence of the binary companion (see Figures 7 --
12).  Specifically, the number of terrestrial planets and the spatial
extent of the terrestrial planet region both decrease with decreasing
binary periastron.  When the periastron value becomes as small as 5
AU, planets no longer form with $a = 1$ AU orbits and the mass
distribution tilts toward $m_p <$ 1 M$_{\oplus}$, i.e., the formation of
Earth-like planets is compromised.  Note that the results from our
simulations can be scaled for different star and disk
parameters with the formulae presented in Appendix C of Quintana \&
Lissauer (2006).  Given the enormous range of orbital parameter space
sampled by known binary systems, from contact binaries to separations
of nearly a parsec, the range of periastron where terrestrial planet
formation is affected is quite similar to the range of periastron
where the stability of Earth-like planets is compromised. As a result,
about 40 -- 50\% of binaries are wide enough to allow both the
formation and the long term stability of Earth-like planets in S-type
orbits encircling one of the stars.  Furthermore, approximately 10\% of
main sequence binaries are close enough to allow the formation and
long-term stability of terrestrial planets in P-type circumbinary
orbits (David \etal\ 2003, Quintana \& Lissauer 2006). Given that the
galaxy contains more than 100 billion star systems, and that roughly half
remain viable for the formation and maintenence of Earth-like planets,
a large number of systems remain habitable based on the dynamic
considerations of this research.

\acknowledgements
We thank Michael J. Way for providing additional CPUs at NASA ARC.
E.V.Q. received support in various stages of this research from NASA
GSRP, NAS/NRC and NASA NPP fellowships, and the University of Michigan
through the Michigan Center for Theoretical Physics (MCTP).  F.C.A. is
supported by the MCTP, and NASA through the Terrestrial Planet Finder
Mission (NNG04G190G) and the Astrophysics Theory Program
(NNG04GK56G0).  J.J.L. is supported in part by the NASA Astrobiology
Institute under the NASA Ames Investigation ``Linking our Origins to
our Future''.

{}

\clearpage
\begin{deluxetable}{ll|lllllll}
\tablecolumns{9}
\tabletypesize{\scriptsize}
\tablecaption{The Outermost Stable S-type Orbit ($a_{c}$)}
\label{tbl-1}
\tablewidth{0pt}
\tablehead{
\colhead{$e_B$} &
\colhead{} &
\colhead{} &
\colhead{$\mu$=1/3} &
\colhead{} &
\colhead{$\mu$=1/2} &
\colhead{} &
\colhead{$\mu$=2/3} &
\colhead{} }

\startdata

0      &  &  & -    & & 0.26 & & -    &\\
0.25   &  &  & 0.23 & & 0.20 & & 0.11 &\\
0.5    &  &  & -    & & 0.12 & & -    &\\
0.625  &  &  & -    & & 0.08 & & -    &\\
0.75   &  &  & -    & & 0.06 & & -    &\\
0.8125 &  &  & -    & & 0.04 & & -    &\\
0.875  &  &  & -    & & 0.03 & & -    &\\
\enddata
\end{deluxetable}
\clearpage

\begin{deluxetable}{lllllll}
\tablecolumns{7}
\tabletypesize{\scriptsize}
\tablecaption{Binary Star Parameters}
\label{tbl-2}
\tablewidth{0pt}
\tablehead{

\colhead{} &
\colhead{} &
\colhead{$q_B$ (AU)} &
\colhead{$a_B$ (AU)} &
\colhead{$e_B$} &
\colhead{$a_{c}$ (AU)} &
\colhead{\# Runs}
}
\startdata

Set A$^{\star}$
& &5 &10 &0.5 &1.2 &5\\
& &5 &20 &0.75 &1.2 &3\\
& &5 &40 &0.875 &1.2 &3\\
& &7.5 &10 &0.25 &2 &5\\
& &7.5 &20 &0.625 &1.6 &3\\
& &7.5 &40 &0.8125 &1.6 &3\\
& &10 &10 &0 &2.6 &5\\
& &10 &13$\frac{1}{3}$&0.25 &2.6 &3\\
& &10 &20 &0.5 &2.4 &3\\
& &10 &40 &0.75 &2.4 &5\\

& & & & & &\\

Set B
& &5 &10 &0.5 &1.2 &10\\
& &7.5 &10 &0.25 &2 &30\\
& &10 &10 &0 &2.6 &10\\
& &10 &40 &0.75 &2.4 &20\\

& & & & & &\\

Set C& &7.5 &10 &0.25 &1.1 &5\\

& & & & & &\\

Set D& &7.5 &10 &0.25 &2.3 &5\\

\enddata
\tablenotetext{\star}{The four sets of accretion simulations are grouped
  according to the masses of the binary stars.  Set A ($M_{\star}$ =
  $M_C$ = 0.5 M$_{\odot}$) and Set B ($M_{\star}$ = $M_C$ = 1
  M$_{\odot}$) have the same stellar mass ratios of $\mu$ = 1/2.  In
  Set C ($M_{\star}$ = 0.5 M$_{\odot}$ and $M_C$ = 1 M$_{\odot}$),
  $\mu$ = 2/3, while Set D ($M_{\star}$ = 1 M$_{\odot}$ and $M_C$ = 0.5
  M$_{\odot}$) has $\mu$ = 1/3.  Herein, $M_{\star}$ indicates the
  `primary' star around which the disk is centered, and $M_C$ is the
  stellar companion. }

 \end{deluxetable}

\clearpage

\begin{deluxetable}{lllllllllllllll}
\tablecolumns{13} 
\tabletypesize{\scriptsize} 
\rotate
\tablecaption{Statistics for Final Planetary Systems}
\label{tbl-3}
\tablewidth{0pt} 
\tablehead{
\colhead{$\bf System$} &
\colhead{\bf \# Runs} &
\colhead{\bf $N_p$}   &
\colhead{\bf $N_m$}   &
\colhead{\bf $a_{p_{max}}$}   &
\colhead{\bf $a_{p_{max}}$/$a_{c}$}   &
\colhead{\bf $Q_p$}   &
\colhead{\bf $Q_p$/$q_B$}   &
\colhead{$S_m$}     & 
\colhead{\bf $m_{l_\star}$} & 
\colhead{\bf $m_{l_\infty}$} & 
\colhead{\bf $E/E_{0}$} & 
\colhead{$L/L_{0}$}}

\startdata 

A\_10\_.5    &5&2   & 0.2  & 0.90 & 0.65 & 0.95& 0.17  &  0.66&  34.9 & 31.5 & 1.64 &  0.74 \\

A\_20\_.75   &3& 1.3& 0&  0.71  & 0.68&  0.52 &   0.13&        0.93&        40.4& 31.3& 1.76 &   0.71\\
%

A\_40\_.875  &3&1   &    0    &0.60 &  0.47&0.52 &       0.13&        1.00&       42.9& 34.4& 1.86& 0.67\\

%

A\_10\_.25&5& 3.2   &  0     &1.32 & 0.59  & 1.43& 0.17  & 0.50 &  9.6 &  39.8 & 1.49  & 0.79\\
A\_20\_.625&3& 2.3& 0 &1.32&      0.79&1.24 &       0.18&    0.68&  38.9& 8.8& 1.43& 0.81\\

%
A\_40\_.8125&3&  1.7& 0.3&0.97&        0.58&0.85 &       0.14 &   0.64&   34.2& 18.1& 1.52& 0.78\\
%

A\_10\_0&5& 3.2 &    1.2 & 2.07  & 0.77 &2.50 &  0.23 &  0.50&  0.2  & 39.5 &   1.39&   0.84\\
A\_13.3\_.25 &3&3&       0.3&1.68 &      0.50&1.50&        0.14&        0.41&       5.36& 27.0& 1.31& 0.85\\

%
A\_20\_.5 &3& 2.3& 0&   1.56 &   0.59&1.39  &  0.16&        0.52&       30.8& 1.7& 1.31& 0.87\\
%
A\_40\_.75  &5& 1.8  &   0     &1.47&  0.45 &2.06&  0.13 &  0.68  & 26.4 & 12.4 & 1.44&   0.81 \\

B\_10\_.5 &10&1.8    & 0.6   & 0.87 &0.62  &0.99&  0.16 &  0.73& 35.5  &29.2& 1.62 &  0.74 \\


B\_10\_.25  &30& 2.8  &   0.5   &1.65  & 0.58  &1.81& 0.17    &    0.56& 9.8& 30.1 & 1.36& 0.83  \\

B\_10\_0  &10&3.8    & 0.7    &2.15& 0.64 &2.78 & 0.18  & 0.47  & 0.0     &  25.7 & 1.22&   0.90 \\
B\_40\_.75  &20&2.6  &  0.2  &1.85&  0.55 &1.99&  0.15 & 0.60&  19.2 &  7.5 & 1.26 & 0.87 \\

C\_10\_.25\_2/3 &5&2.8  &   0.6  &1.05&   0.81 &1.06&  0.12 &  0.53& 0.0     &  72.4&  1.66  & 0.74 \\

D\_10\_.25\_1/3 & 5&3.4   &  0.2    &1.58& 0.61  &1.79& 0.21 &  0.54&25.5&  0.3 &  1.24 & 0.88 \\

  & & & & & & & & & & & & &   \\
MVEM     &$\cdots$&4.0&0.0&&$\cdots$&&$\cdots$&0.51&$\cdots$&$\cdots$&$\cdots$&$\cdots$\\    
SJS &31&3.0&0.7&&$\cdots$&&$\cdots$&0.51&23.9&2.0&1.26&0.87\\    
Sun  &3&4.3&12&&$\cdots$&&$\cdots$&0.39&0.0&0.5&1.06&1.01\\    
Sun ($a <$ 2 AU)  &3&3.0&0.7&&$\cdots$&&$\cdots$&0.48&0.0&18.7$^\ddag$ &1.21&0.90\\    
$\alpha$ Cen A ($i$ = 0$^{\circ}$)&4&4.3&0&&$\cdots$&&$\cdots$&0.40&11.6&0.8&1.12&0.94\\

\enddata

\tablenotetext{\ddag}{Mass lost plus mass ending up in planets/minor planets beyond 2 AU.}

\end{deluxetable} 

\clearpage

\begin{figure}
\figurenum{1}
\epsscale{0.80} 
\plotone{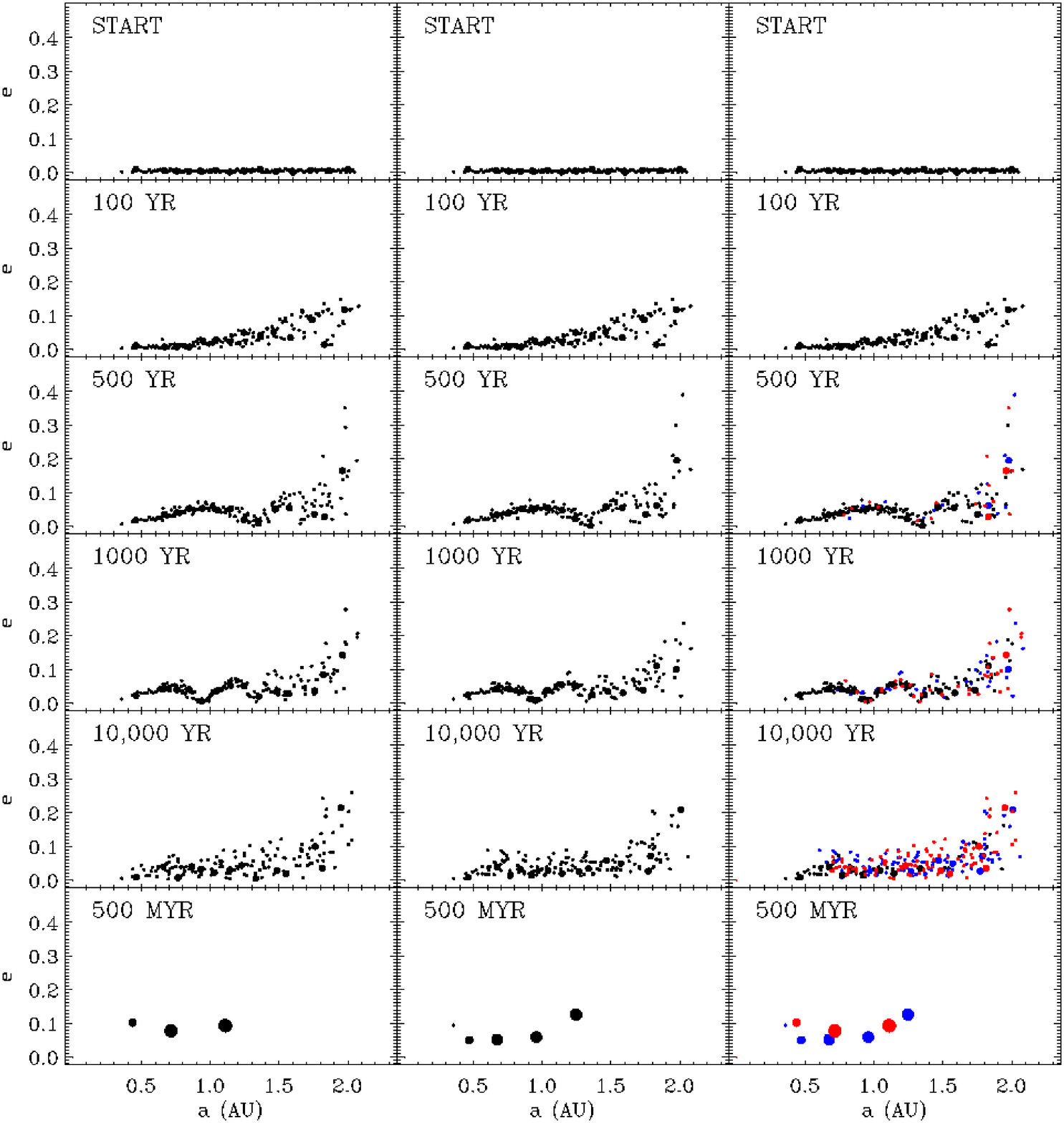}
\figcaption{\small{The temporal evolution of two simulations, each with $a_B$
  = 10 AU, $e_B$ = 0.25, and equal mass stars of 1 M$_{\odot}$, is
  shown in the first two columns.  The only difference between these
  two simulations at the start of the integration is that one
  planetesimal near 1 AU (in the second run) is shifted forward by 1
  meter along its orbit.  The third column overlays the two systems at
  the corresponding time in order to demonstrate the divergence of
  initially nearby orbits.  The bodies in the disk are represented by
  circles whose sizes are proportional to the physical sizes of the
  bodies, and whose locations show the orbital semimajor axes and
  eccentricities of the represented bodies relative to the central
  star.  In the third column, bodies for which $(e_{p_{1}} - e_{p_{2}})^2 + (a_{p_{1}} -
  a_{p_{2}})^2 >$ 0.001, where $e_p$ and $a_p$ are the eccentricity and
  semimajor axis of each planetesimal/embyro, are plotted in red (for
  the bodies from the first column) and in blue (for the corresponding
  bodies in the second column).  Note that while these systems seem
  qualitatively similar at early times, the two systems begin to
  diverge within 500 years of the simulation.  The small change in the
  initial conditions ultimately leads to the formation of two very
  different planetary systems, 3 planets versus 5, as shown in the
  final row. }}
\end{figure}

\begin{figure}
\figurenum{2}
\centering
\includegraphics[width=3.7in, angle=270]{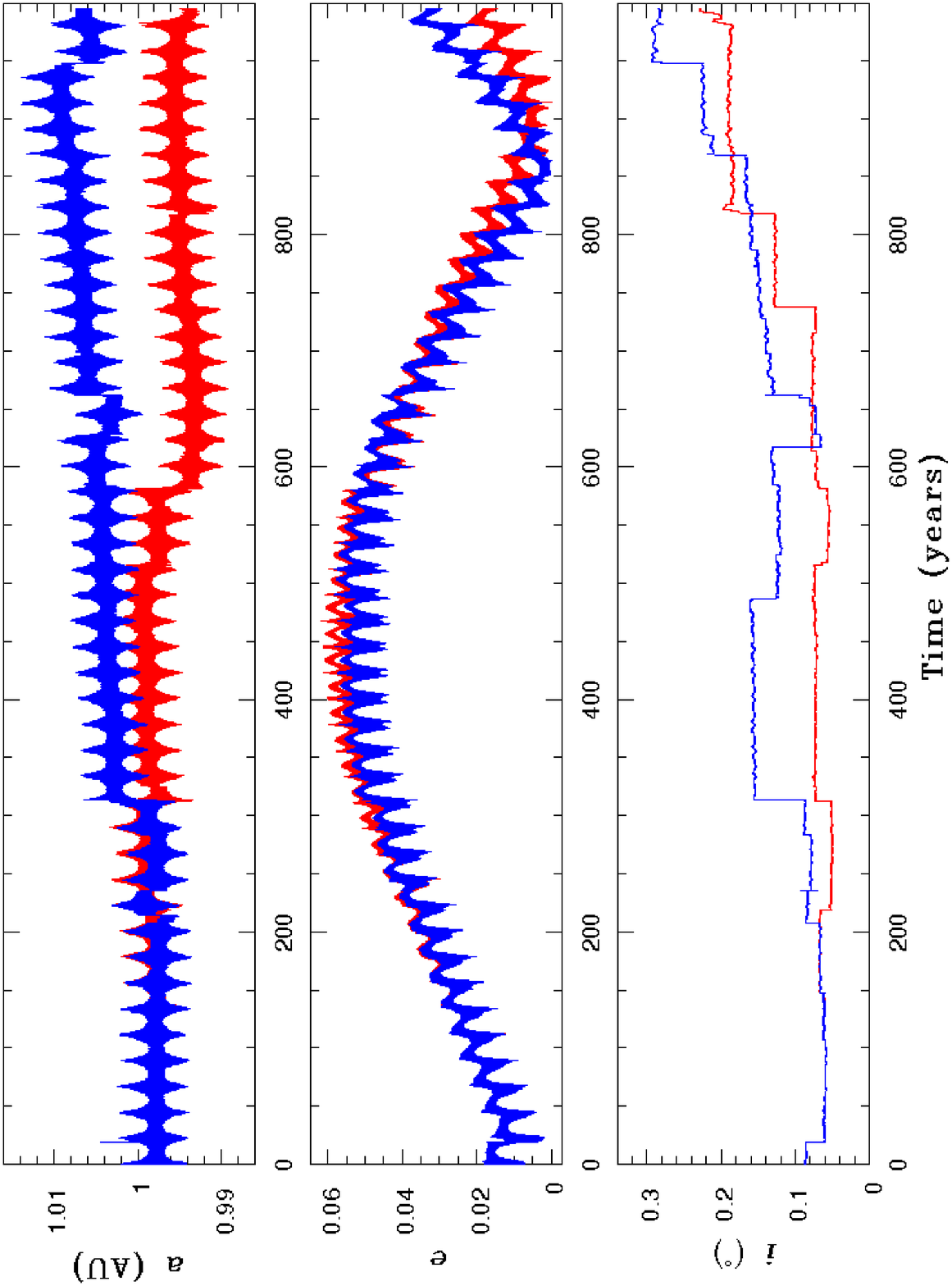}
\includegraphics[width=3.9in, angle=270]{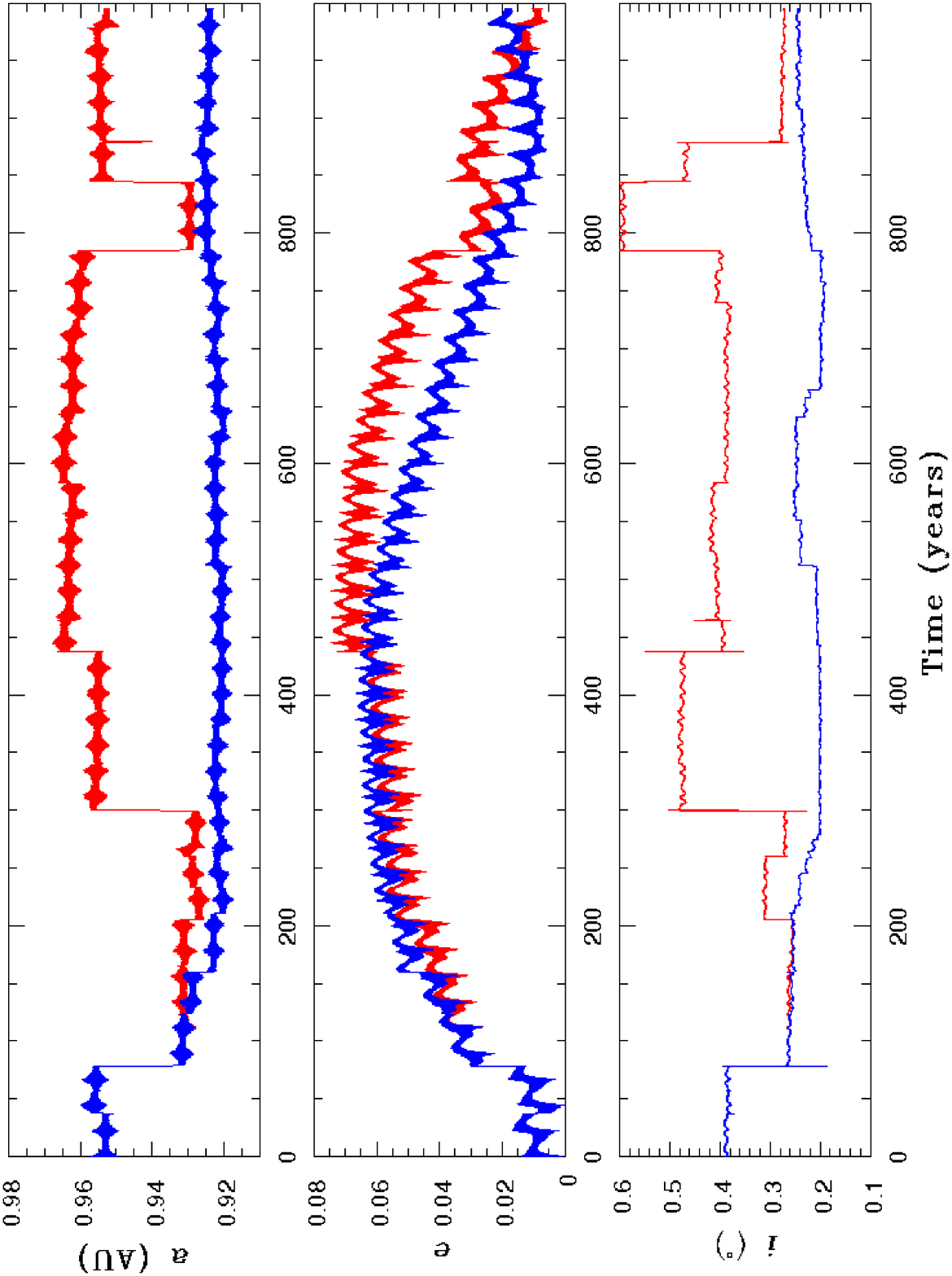}
\figcaption{\small{The time evolution of the semimajor axis ($a_p$),
  eccentricity ($e_p$), and inclination ($i_p$) is shown for a
  planetesimal on an initial orbit near 1 AU (upper 3 panels), and for another
  planetesimal that began near 0.95 AU (lower 3 panels), from the first
  1000 years of the two simulations shown in Figure 1.  (a) In the
  top figure, the initial orbits of the two planetesimals differ by 1 meter in their
  mean anomoly, while all other system parameters are identical.  (b) In the bottom figure, the two planetesimals
  began on identical orbits near 0.95 AU (also shown in red and blue).
  The chaotic behavior of these systems is apparent early in the
  simulations, and the orbital elements of nearby trajectories diverge
  with Lyapunov times of $\sim$ 100 -- 200 years.}}
\end{figure}
 
\begin{figure} 
\centering
\figurenum{3} \epsscale{0.80}
\plotone{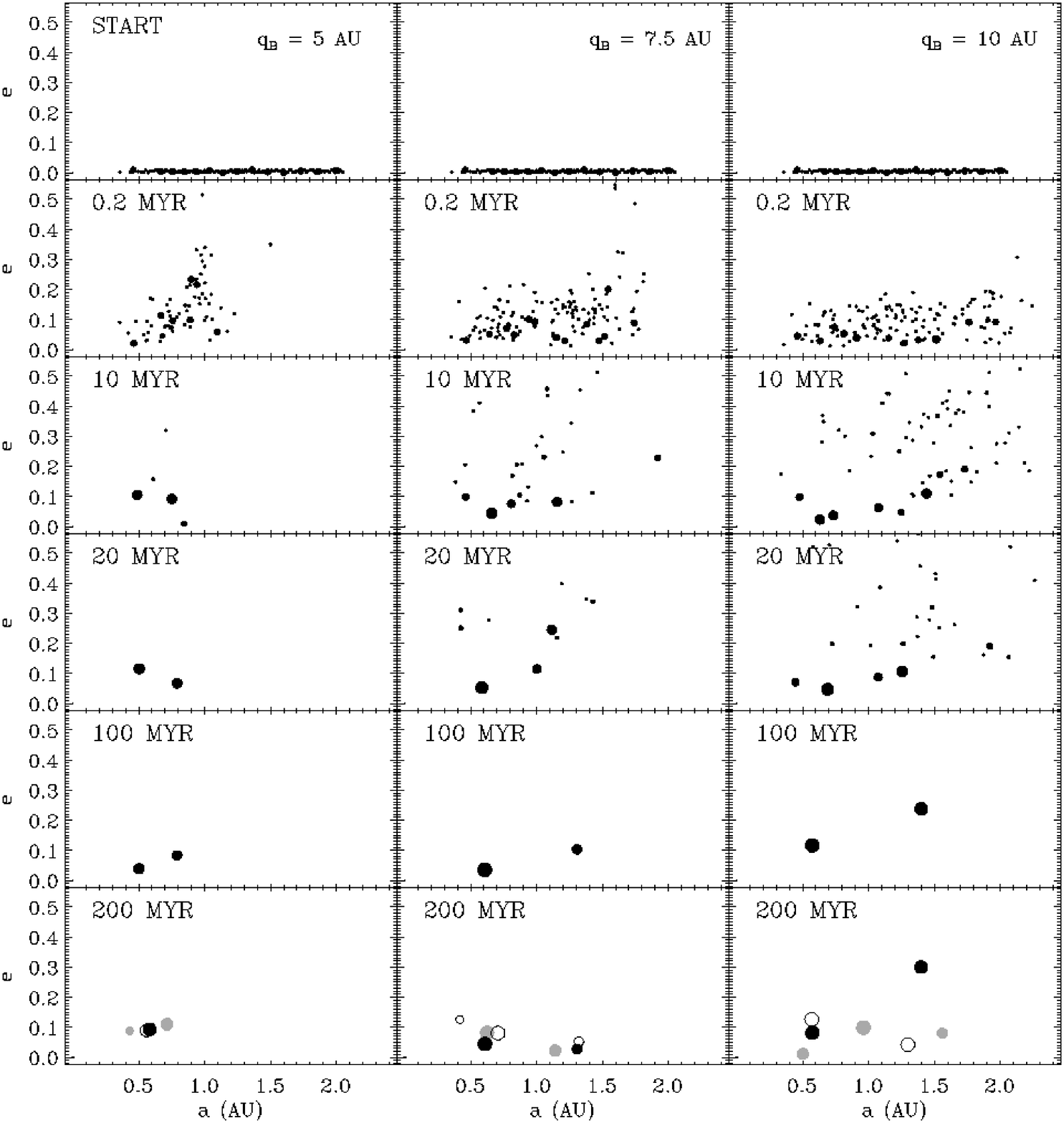}
\figcaption{\small{The temporal evolution of three systems from Set A is
  shown here, with each body in the disk represented by black solid
  circles (with symbol sizes proportional to planet sizes).  These
  three simulations each begin with equal mass stars of 0.5
  M$_{\odot}$ and semimajor axis $a_B$ = 20 AU.  The simulations
  differ in their stellar eccentricities (and therefore periastra):
  $e_B$ = 0.75 and $q_B$ = 5 AU (first column), $e_B$ = 0.625 and
  $q_B$ = 7.5 AU (middle column), and $e_B$ = 0.5 and $q_B$ = 10 AU
  (third column).  The disk is clearly truncated in the $q_B$ = 5 AU
  and 7 AU simulations, but still dynamically excited in each case.
  The amount of mass that is lost (averaged for each system) is approximately
  71\% ($q_B$ = 5 AU), 48\% ($q_B$ = 7.5 AU), and 33\% ($q_B$ = 10
  AU).  The final planets formed from two additional realizations of
  each system (with the same stellar parameters) are shown in the last
  row with gray and open circles; their early evolution (not shown) has
  similar trends as the simulation shown with black solid symbols.}}
\end{figure}

\begin{figure}
\centering
\figurenum{4} \epsscale{0.80} 
\plotone{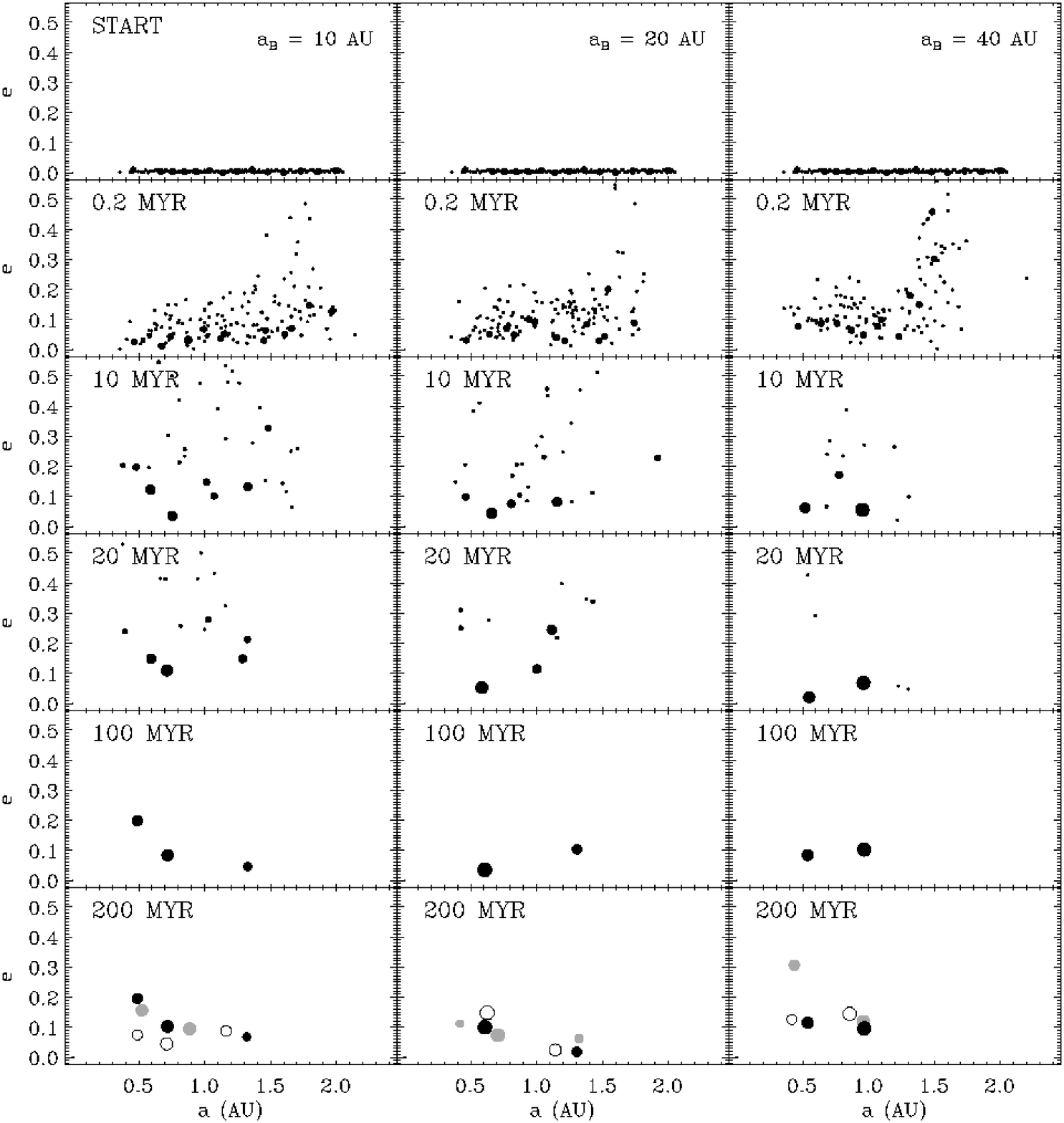}
\figcaption{\small{The temporal evolution of the bodies in the disk from
  three simulations of Set A ($M_{\star}$ = $M_C$ = 0.5 M$_{\odot}$),
  each of which began with a binary periastron of $q_B$ = 7.5 AU, are
  represented in each column with black filled circles.  The
  simulations differ in the stellar semimajor axes $a_B$ and
  eccentricities $e_B$.  In the first column, $a_B$ = 10 AU and $e_B$
  = 0.25, in the middle column $a_B$ = 20 AU and $e_B$ = 0.625, and in
  the third column $a_B$ = 40 AU and $e_B$ = 0.8125.  The bottom row
  displays the results of two additional simulations (shown by gray and open circles) which began with the same stellar parameters.  See Figure 3
  for an explanation of the symbols.  Early in each simulation, many
  of the planetesimals near the outer edge of the disk are lost from
  the system, as are the more massive embryos in simulations that
  begin with larger $a_B$ and $e_B$.  The total mass loss from the
  three integrations is comparable, $\sim$ 50\%, although most of the
  mass that is lost from the more eccentric binary star systems (middle and third columns) typically ends up being removed by passing too close (within 0.1 AU) to the central star. }}
\end{figure}

\begin{figure}
\centering 
\figurenum{5} \epsscale{0.80}
\plotone{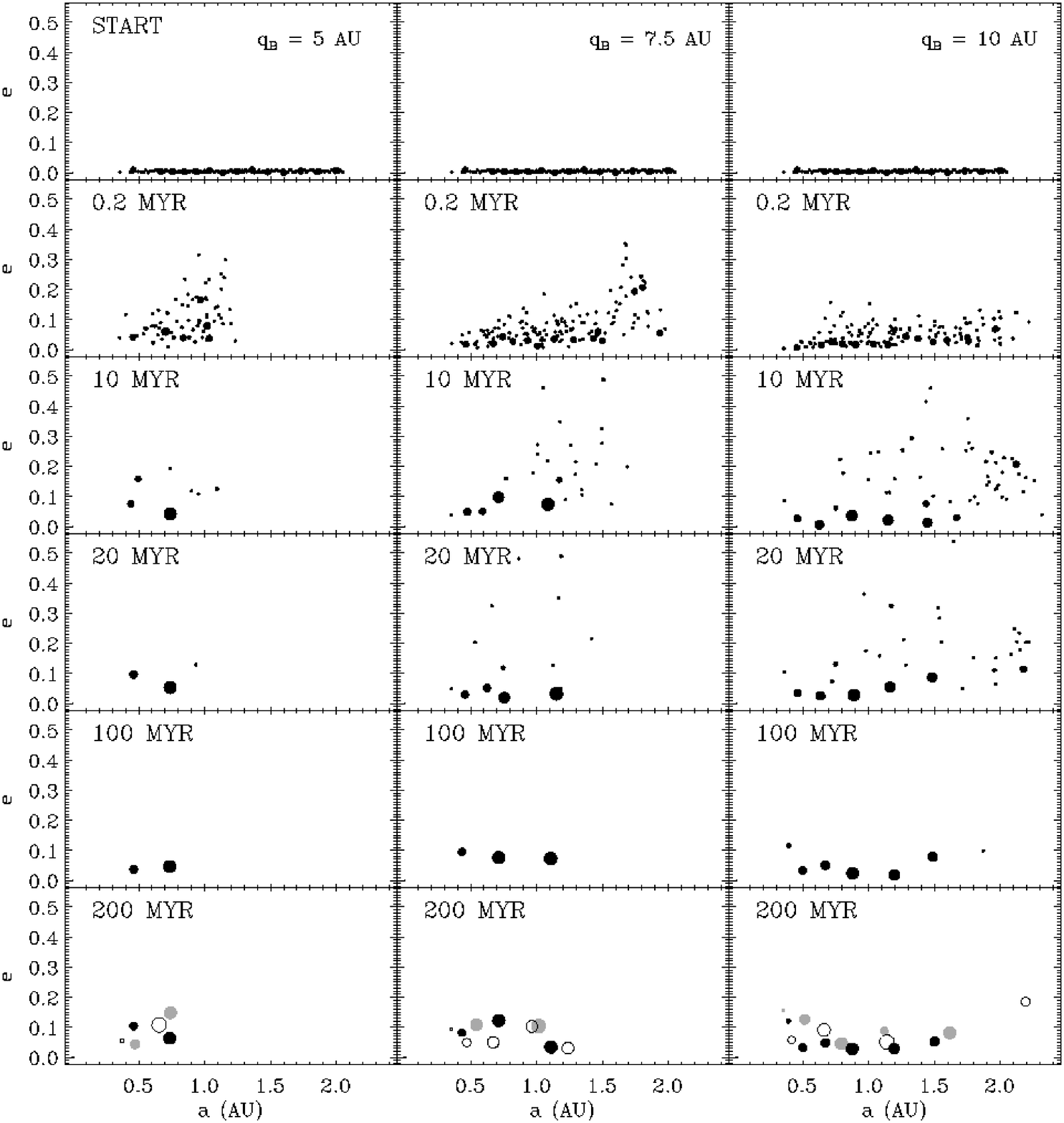}
\figcaption{\small{The temporal evolution of the bodies in the disk from
  three simulations of Set B ($M_{\star}$ = $M_C$ = 1 M$_{\odot}$),
  each of which begin with $a_B$ = 10 AU, is shown in each column with
  black filled circles.  These simulations differ in the binary
  eccentricity and periastra: $e_B$ = 0.5 and $q_B$ = 5 AU (left
  column), $e_B$ = 0.25 and $q_B$ = 7.5 AU (middle column), and $e_B$
  = 0 and $q_B$ = 10 AU (right column).  The bottom row also displays
  the results of two additional simulations (shown by gray and open circles) using
  the same stellar parameters.  See Figure 3 for the explanation of
  the symbols.  These simulations display similar accretion time
  scales, and result in similar final planetary systems, as analogous
  simulations from Set A which have the same stellar mass ratio $\mu$
  = 0.5, but consist of equal mass stars of 0.5 M$_{\odot}$.  The bodies in the disk
  in Set B, however, are slightly less perturbed than those in Set A due to the smaller mass ratio of the planetesimals/embryos to the stars, which leads to systems with slightly larger radial extents than those formed in Set A.}}
\end{figure}

\begin{figure}
\centering \figurenum{6} \epsscale{0.80}
\plotone{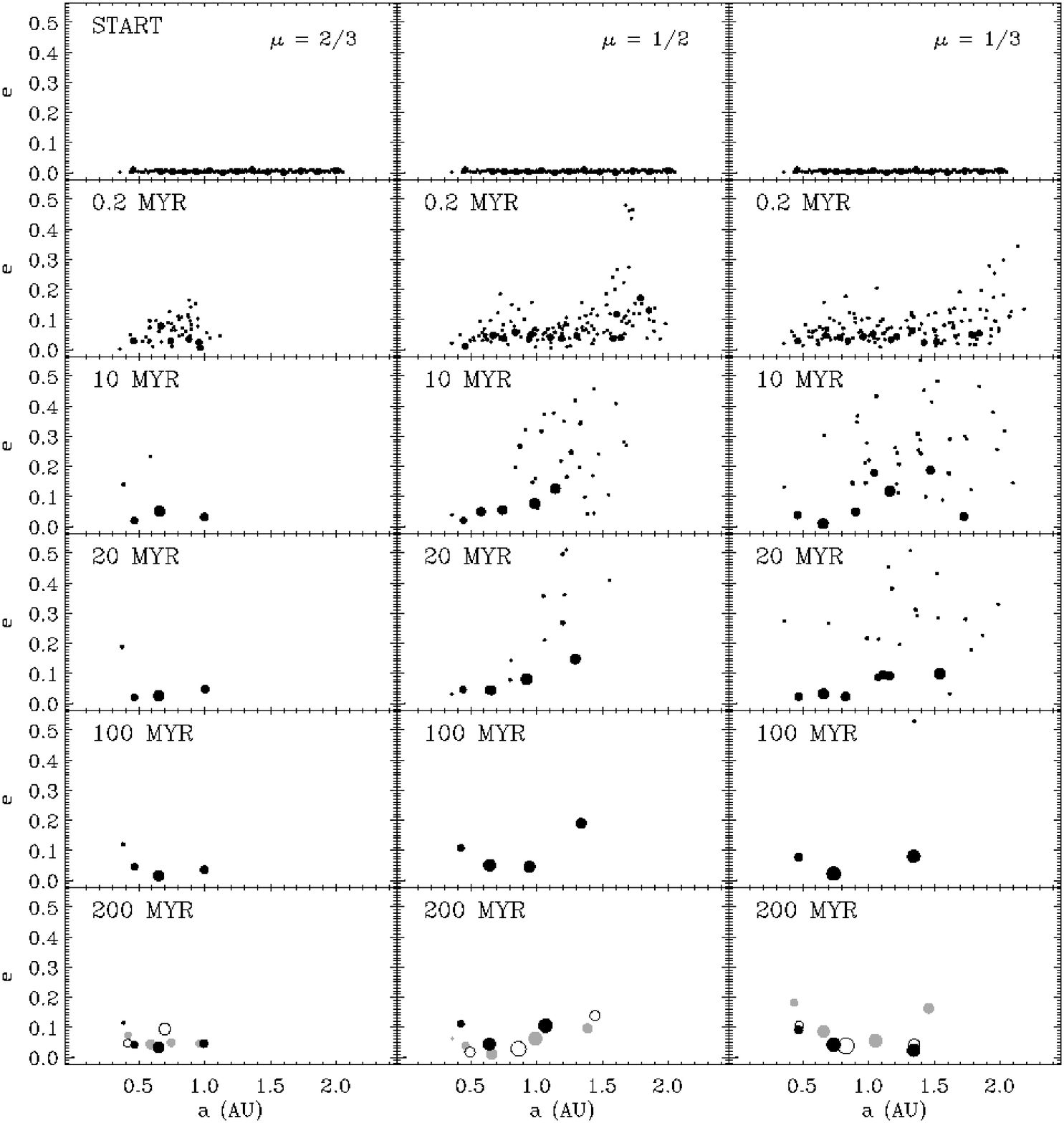} 
\figcaption{\small{The temporal
  evolution of three simulations (shown in each column by black filled
  circles) with different stellar mass ratios, but the same orbital
  parameters: $q_B$ = 7.5 AU, $a_B$ = 10 AU, and $e_B$ = 0.25.  The
  left column shows a system with $\mu$ = 2/3 ($M_{\star}$ = 0.5
  M$_{\odot}$ and $M_C$ = 1 M$_{\odot}$), the system in the middle
  column has $\mu$ = 1/2 ($M_{\star}$ = $M_C$ = 1 M$_{\odot}$), and
  the third column shows $\mu$ = 1/3 (in which $M_{\star}$ = 1
  M$_{\odot}$ and $M_C$ = 0.5 M$_{\odot}$).  The bottom row also
  displays the results of two additional simulations (shown by gray and open circles) using the same stellar parameters.  See Figure 3 for the
  explanation of the symbols.  Although the amount of mass that is
  lost in these systems is quite different (an average of 72\%, 40\%,
  and 25\% for columns 1, 2, and 3, respectively), the fraction of
  mass in the largest planet within each system is comparable,
  $\sim$ 50\% ($S_m \sim$ 0.5) on average.}}
\end{figure}
\clearpage
\thispagestyle{empty}
\begin{figure} 
\figurenum{7} 
\centering
\vspace*{-14mm}
\includegraphics[width=5in]{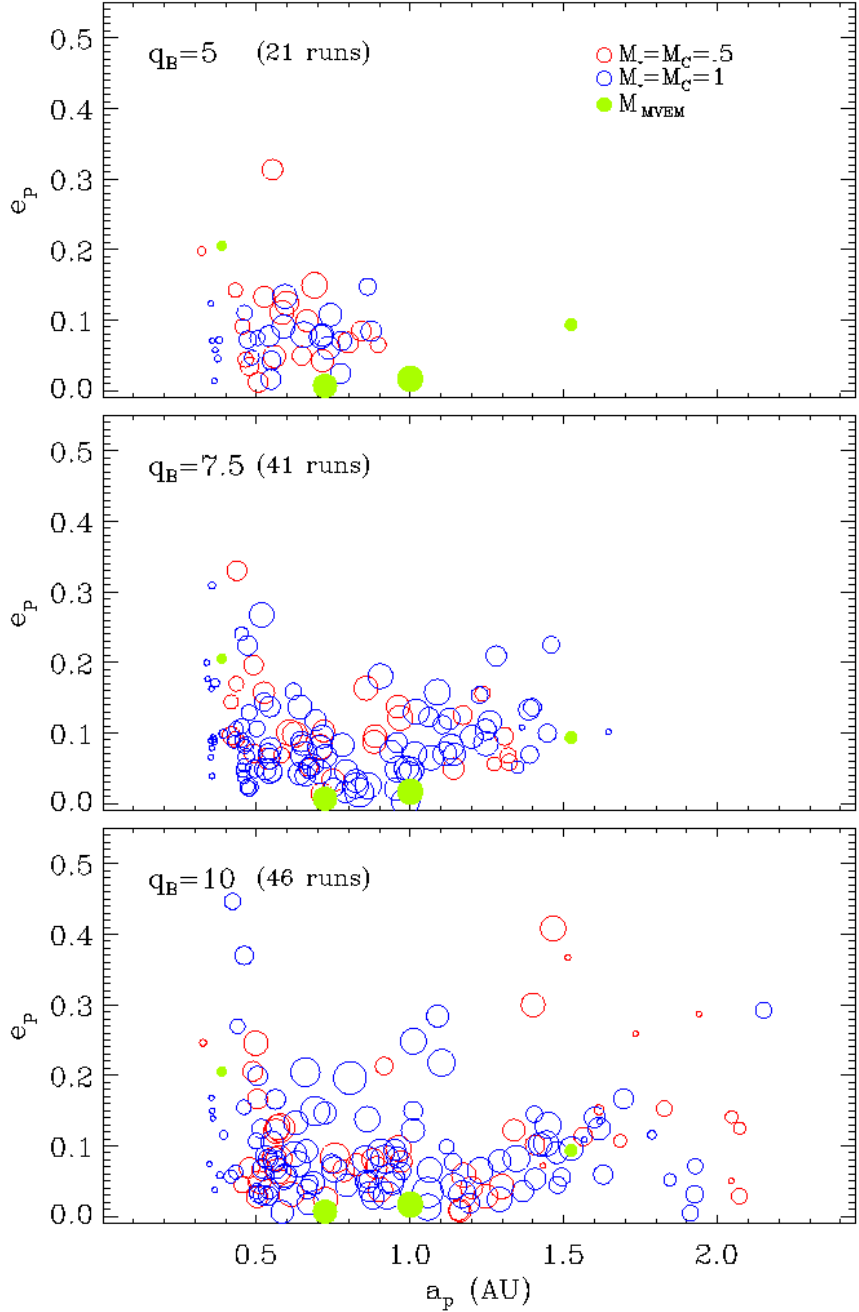} 
\figcaption{\small{The eccentricity as a function of semimajor axis is shown
  for all of the final planets formed in equal-mass binary star
  systems with $q_B$ = 5 AU (top panel), $q_B$ = 7.5 AU (middle
  panel), and $q_B$ = 10 AU (bottom panel).  The radius of each symbol
  is proportional to the body that it represents, and the terrestrial
  planets in our Solar System (at the J2000 epoch) are represented by
  green filled circles in each panel for comparison.  The red symbols
  represent binary systems from Set A with $M_{\star}$ = $M_C$ = 0.5 M$_{\odot}$,
  while the blue symbols represent systems from Set B with $M_{\star}$ = $M_C$ =
  1.0 M$_{\odot}$. The two distributions for each $q_B$ are
  comparable, although planetary perturbations in the Set B
  simulations are relatively smaller than those of Set A due to the
  smaller ratio of planet-to-star mass, which results in less internal
  excitation.  This effect is evident in the larger number of
  planetesimals that have remained close to the central star in the
  Set B simulations shown in each panel.}}
\end{figure}
\clearpage 
\begin{figure}
\figurenum{8} 
\centering
\includegraphics[angle=270, width=18cm]{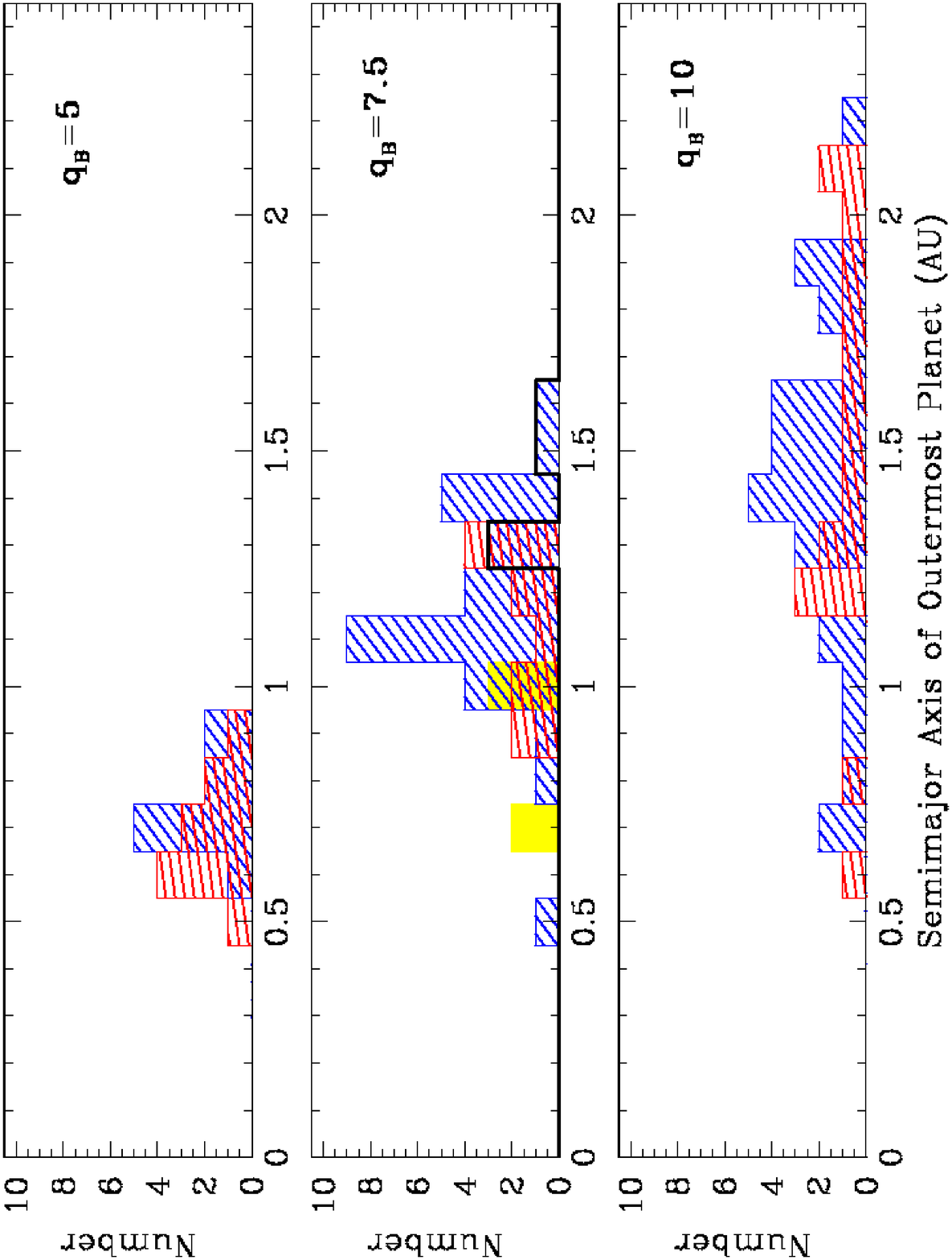} 
\figcaption{\small{The distribution of the semimajor axis of the
    outermost final planet formed for binary star systems with $q_B$ =
    5 AU (top panel), $q_B$ = 7.5 AU (middle panel), and $q_B$ = 10 AU
    (bottom panel).  The red bars represent simulations from Set A
    with $M_{\star}$ = $M_C$ = 0.5 M$_{\odot}$, whereas the blue bars
    represent systems from Set B with $M_{\star}$ = $M_C$ = 1.0
    M$_{\odot}$.  Also shown in the middle panel are the results from
    Set C in which $M_{\star}$ = 0.5 M$_{\odot}$ and $M_C$ = 1.0
    M$_{\odot}$ (solid yellow bars), and from Set D with $M_{\star}$ =
    1.0 M$_{\odot}$ and $M_C$ = 0.5 M$_{\odot}$ (solid black line).
    Although the semimajor axes extend to larger values in binary
    systems with larger periastron, the inner edge of the distribution
    is roughly determined by the inner edge of the initial disk of
    embryos, as in the single star case, i.e., the presence of
    different stellar companions has a minimal effect on the inner
    terrestrial region.}}
\end{figure}

\begin{figure}  
\figurenum{9} 
\centering
\includegraphics[width=6.5in]{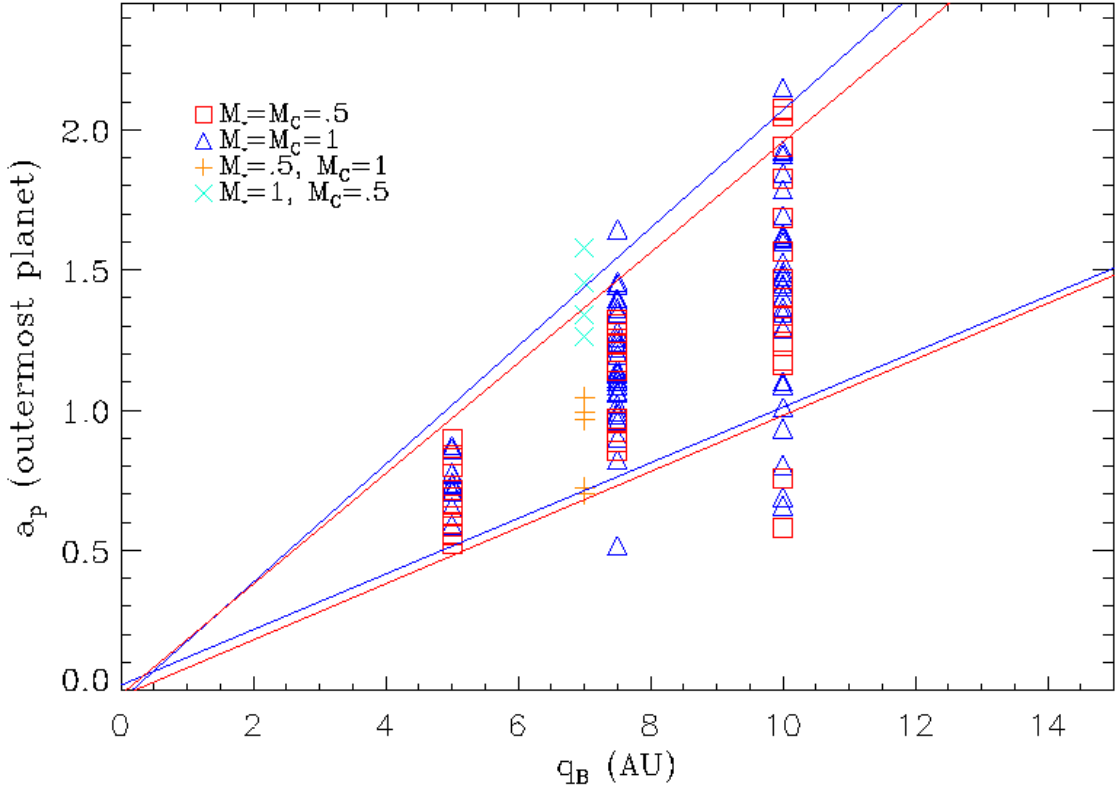}
\figcaption{\small{The semimajor axis of the outermost final planet formed in
  each simulation as a function of binary periastron, $q_B$.  The red
  square symbols represent binary systems with $M_{\star}$ = $M_C$ = 0.5
  M$_{\odot}$ (Set A), while the blue triangle symbols represent systems with
  $M_{\star}$ = $M_C$ = 1.0 M$_{\odot}$ (Set B). A fit to the
  (2$\sigma$) standard deviation for both Sets A and B are also shown
  in red and in blue.  The values from Set C (where $\mu$ = 2/3) are
  shown with orange plus symbols, and those from Set D ($\mu$ = 1/3) are displayed with light blue X symbols. Note that the symbols from Sets C and D have been offset by
  --0.5 AU for clarity.  The expectation value, and also the width of each distribution, grow
in a nearly linear fashion with increasing values of binary
periastron $q_B$.}}
\end{figure}

\begin{figure} 
\figurenum{10} 
\centering
\includegraphics[angle=270, width=18cm]{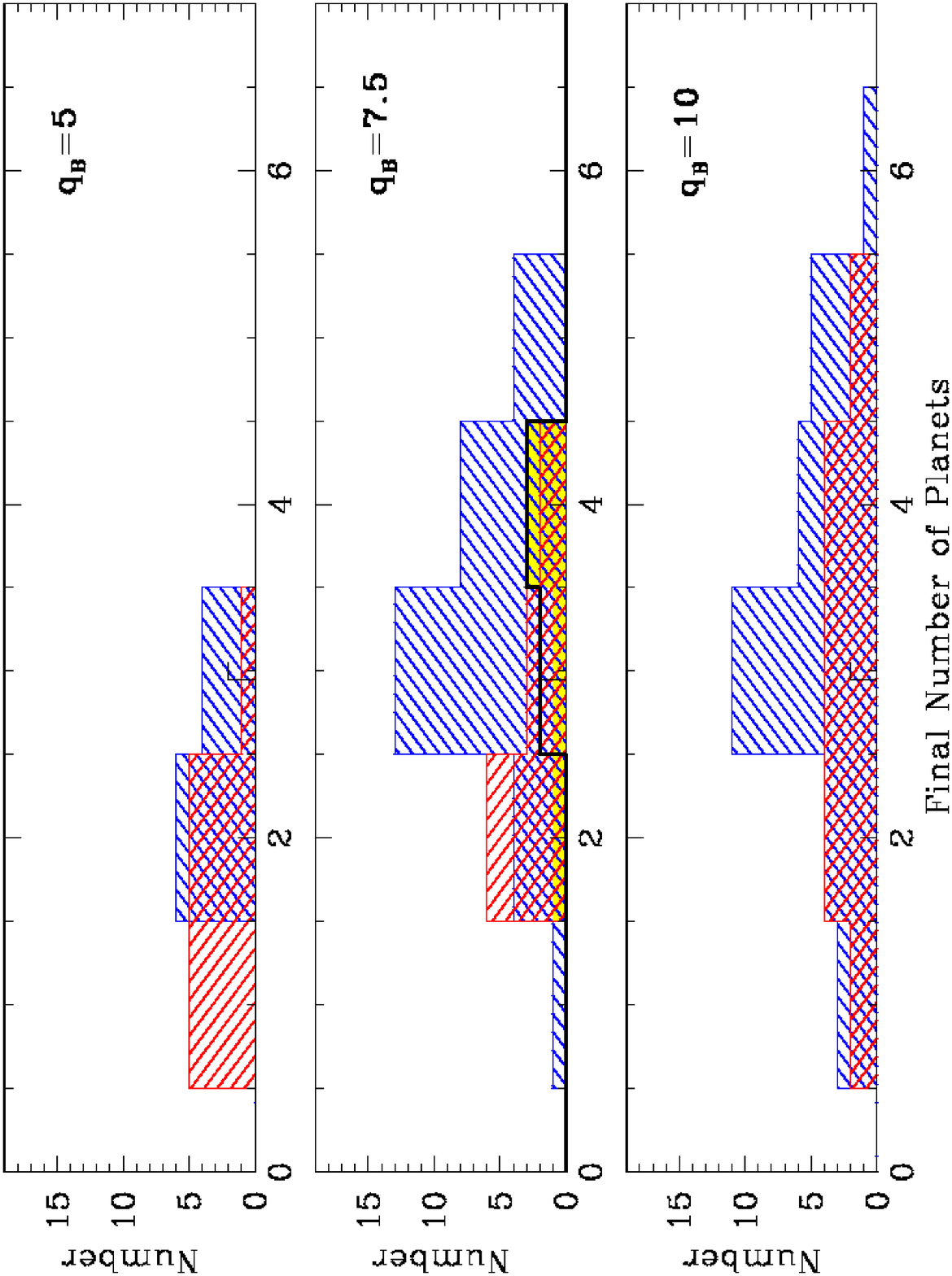}
\figcaption{\small{Histograms of the number of final planets formed for
  binary star systems with $q_B$ = 5 AU (top panel), $q_B$ = 7.5 AU
  (middle panel), and $q_B$ = 10 AU (bottom panel).  The colors correspond
  to the different sets of runs as described in Figure 8.  The
  typical number of final planets clearly increases in systems with
  larger stellar periastron, and also when the companion star is less
  massive than the primary (for a given stellar mass ratio).}}
\end{figure}
  
\begin{figure} 
\figurenum{11} 
\centering
\includegraphics[angle=270, width=18cm]{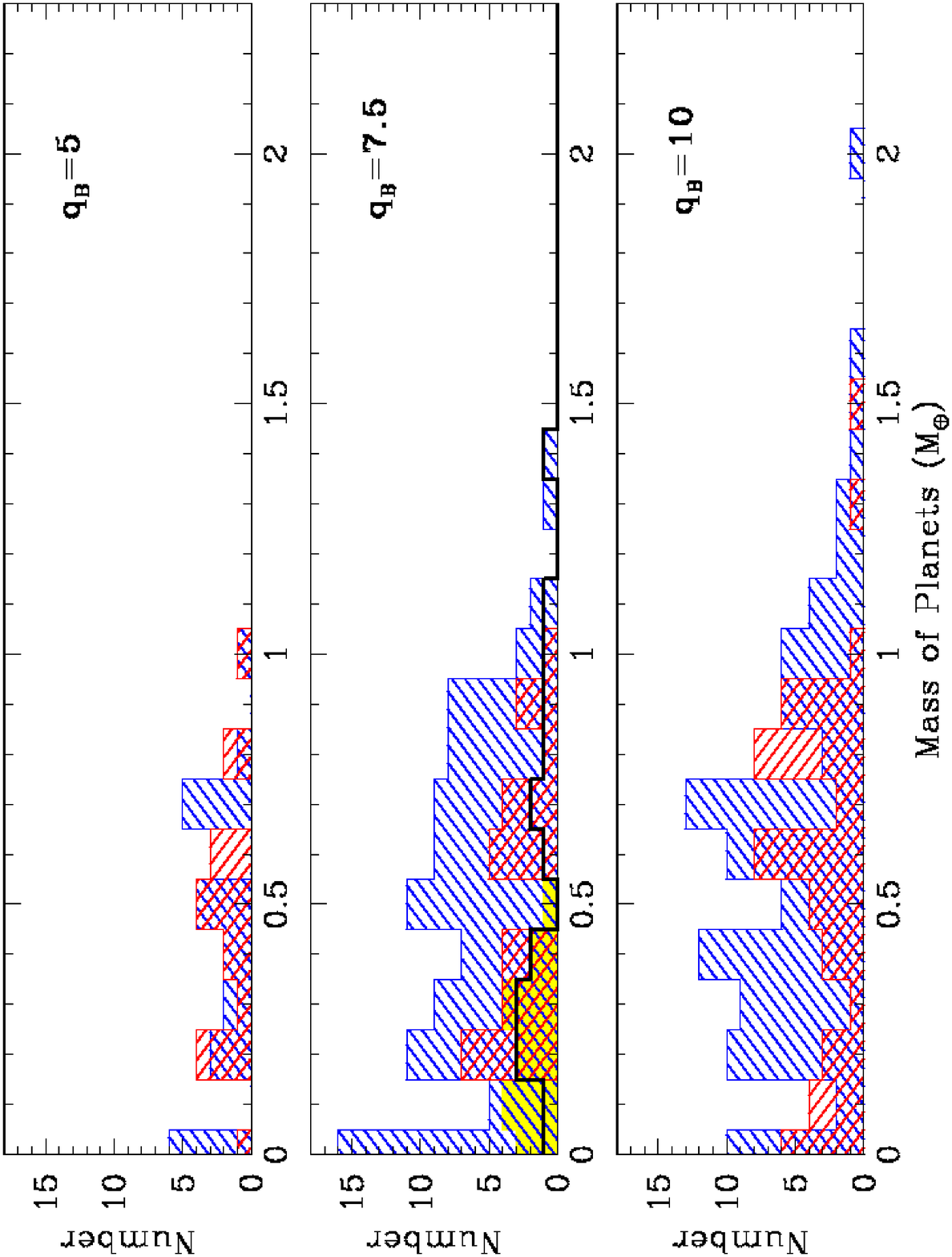}
\figcaption{\small{Histograms of the final masses of planets formed within
  binary star systems with $q_B$ = 5 AU (top panel), $q_B$ = 7.5 AU
  (middle panel), and $q_B$ = 10 AU (bottom panel).  The colors correspond
  to the different sets of runs as described in Figure 8.  Although the
  size of the stable region shrinks as $q_B$ gets smaller, the median
  mass of the final planets does not vary greatly for a given $q_B$,
  suggesting that planet formation remains efficient in the
  stable regions.}}
\end{figure}

\begin{figure}
\figurenum{12} 
\centering
\includegraphics[angle=270, width=18cm]{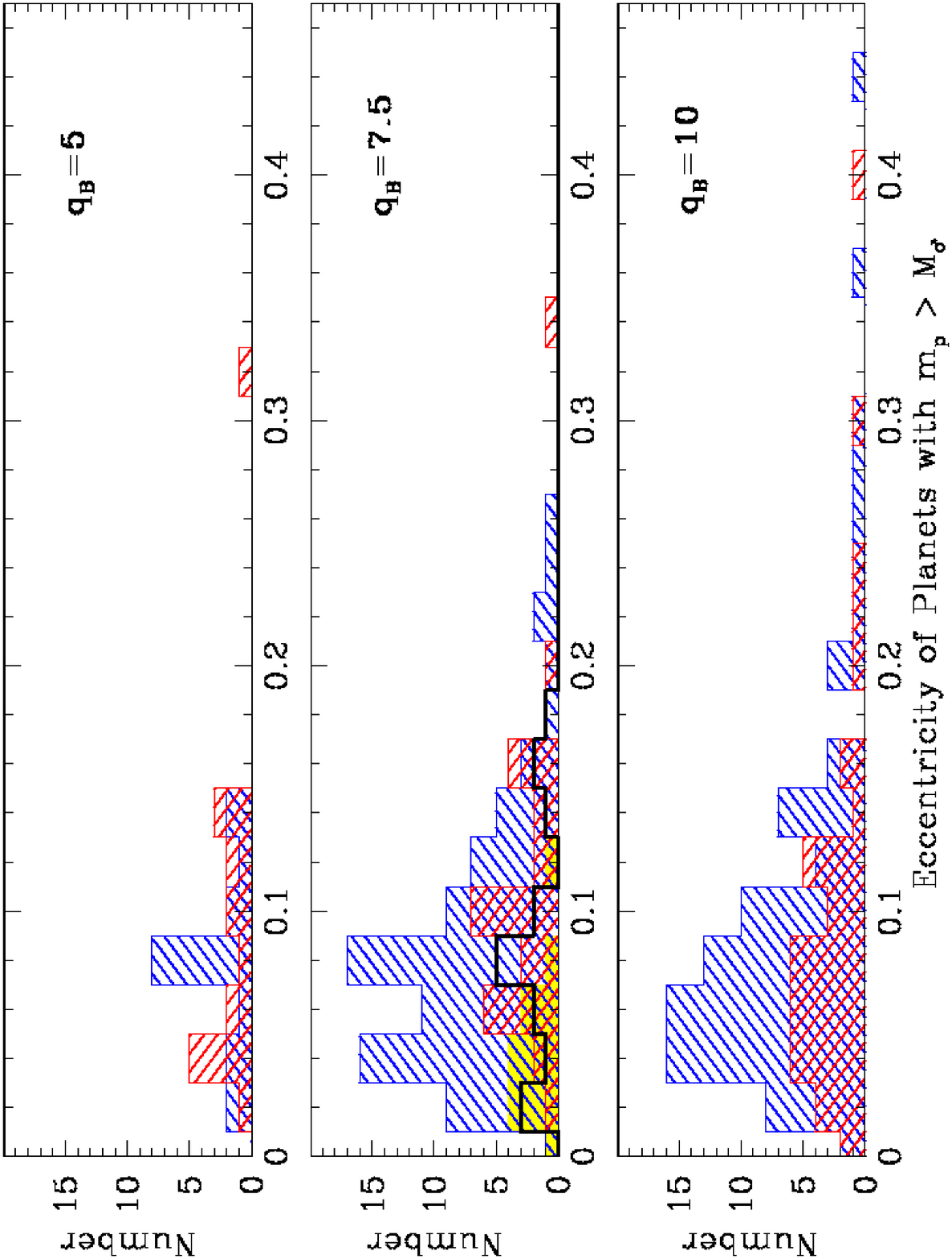}
\figcaption{\small{Distributions of the eccentricities of final planets that
  are more massive than Mars are shown for systems with $q_B$ = 5 AU
  (top panel), $q_B$ = 7.5 AU (middle panel), and $q_B$ = 10 AU
  (bottom panel).  The colors correspond to different sets of runs as
  described in Figure 8.  The majority of planets orbit on
nearly circular ($e \la$ 0.1) orbits despite the proximity of the stellar companion.} }
\end{figure}

\end{document}